\documentclass[a4paper,11pt,reqno]{article}
\usepackage{a4wide}
\usepackage{amsmath,amsfonts,amssymb}
\usepackage[english]{babel}
\usepackage{soul}
\usepackage{nicefrac}
\usepackage[mathscr]{euscript}
\usepackage{setspace}
\usepackage{datetime}
\usepackage[nosort]{cite}
\usepackage{mathrsfs}
\usepackage[T1]{fontenc}
\usepackage{txfonts}
\usepackage{kpfonts}
\usepackage{upgreek}
\usepackage{comment}
\usepackage{mathpazo}
\usepackage{graphicx}
\usepackage[breaklinks=true]{hyperref}
\newcommand{\beq}{\begin{equation}}
\newcommand{\eeq}{\end{equation}}
\newcommand{\beqn}{\begin{eqnarray}}
\newcommand{\eeqn}{\end{eqnarray}}
\newcommand{\pa}{\partial}
\numberwithin{equation}{section}

\hypersetup{colorlinks=true,
			urlcolor=blue,
			citecolor=magenta,
			linkcolor=blue,}

\let\oldsqrt\sqrt
\def\sqrt{\mathpalette\DHLhksqrt}
\def\DHLhksqrt#1#2{%
\setbox0=\hbox{$#1\oldsqrt{#2\,}$}\dimen0=\ht0
\advance\dimen0-0.2\ht0
\setbox2=\hbox{\vrule height\ht0 depth -\dimen0}%
{\box0\lower0.4pt\box2}}

\newcommand{\RNum}[1]{\uppercase\expandafter{\romannumeral #1\relax}}

\author{
  \begin{minipage}{.97\linewidth}
    \vspace{1cm}
       \begin{center}
      \begin{small}
               \textbf{Luca Ciambelli},$^1$ 
             \textbf{Charles Marteau},$^2$ 
                     \\
     \textbf{P. Marios Petropoulos}$^2$ and 
      \textbf{Romain Ruzziconi}$^1$
              \end{small}
    \end{center}
    \vspace{0.5cm}
      \hspace{2.4cm}\begin{minipage}{.7\linewidth}
\begin{center}     {\it \begin{footnotesize}
\hbox{
\kern-1.8cm
\vbox{
\begin{itemize}
 \item[$^1$]Universit\'e Libre de Bruxelles\\ 
   \emph{and} International Solvay Institutes\\
CP 231, 1050 Brussels, Belgium
      \end{itemize}
      \vskip0.35cm
      }   
\kern-3cm
      \vbox{
 \begin{itemize}
                  \item[$^2$]
Centre de Physique Th\'eorique -- CPHT\\ 
        Ecole Polytechnique, CNRS\footnote{\emph{Centre National de la Recherche Scientifique}, Unit\'e Mixte de Recherche UMR 7644.}\\
        Institut Polytechnique de Paris\\
        91128 Palaiseau Cedex, France           
      \end{itemize}}
    }
     \end{footnotesize}}
\end{center}
    \end{minipage}
      \vspace{1.5cm}\begin{minipage}{.7\linewidth}
     \end{minipage}
  \end{minipage}
}

\title{\vspace{2.5cm}
 \boldmath 
    \textbf{
    \textsc{Gauges in Three-Dimensional Gravity and Holographic Fluids}}
  \unboldmath
}

\date{}

\begin{document}

\begin{titlepage}
\maketitle
\thispagestyle{empty}

 \vspace{-13.cm}
  \begin{flushright}
  CPHT-RR021.032020\\
  \end{flushright}
 \vspace{12.cm}

\begin{center}
\textsc{Abstract}\\  
\vspace{1.cm}	
\begin{minipage}{1.0\linewidth}

Solutions to Einstein's vacuum equations in three dimensions are locally maximally symmetric. They are distinguished by their global properties and their investigation often requires a choice of gauge. Although analyses of this sort have been  performed abundantly, several relevant questions remain. These questions include the interplay between the standard Bondi gauge and the Eddington--Finkelstein type of gauge used in the fluid/gravity holographic reconstruction of these spacetimes, as well as  the Fefferman--Graham gauge, when available i.e. in anti de Sitter. The goal of the present work is to set up a thorough dictionary for the available descriptions with emphasis on the  relativistic or Carrollian holographic fluids, which portray the bulk from the boundary in anti-de Sitter or flat instances.  A complete presentation of residual diffeomorphisms with a preliminary study of their algebra accompanies the situations addressed here.

\end{minipage}
\end{center}


\end{titlepage}

\onehalfspace

\begingroup
\hypersetup{linkcolor=black}
\tableofcontents
\endgroup
\noindent\rule{\textwidth}{0.6pt}

\section{Introduction}

Three dimensions have been the playground for investigating various aspects of gravity. Vacuum Einstein spacetimes are locally flat or locally anti-de Sitter (for zero or negative cosmological constant).
Hence, they are characterized by their global properties, encoded in the asymptotic charges. These are computed after specifying a set of boundary conditions, which shape the asymptotic symmetries. In order to implement the boundary conditions and perform the subsequent analyses, it is customary to fix the gauge. Two gauges, called Bondi and Fefferman--Graham, have played a distinguished role.  

The Bondi gauge was introduced in \cite{Bondi1962,Sachs1962,Sachs1962-2} for four-dimensional asymptotically flat spacetimes. It set the stage for the emergence of the celebrated 
Bondi--van der Burg--Metzner--Sachs algebra and recently received revitalized interest from various perspectives \cite{Barnich:2006av,Barnich:2010eb,Barnich:2016lyg},
including the potential implications of this symmetry in infrared physics (see e.g. the review article \cite{Strominger:2017zoo}). The Bondi gauge features a null radial coordinate, and its defining conditions are compatible with asymptotically anti-de Sitter spacetimes. It was extended in this direction in Ref. \cite{Barnich:2013sxa}, while the corresponding flat limit for the three-dimensional case was shown to be consistent in \cite{Glenn3}. In Refs. \cite{Poole:2018koa, Compere:2019bua, Compere:2020lrt} (see also \cite{Ruzziconi:2019pzd}) less stringent conditions were considered, allowing for a unified formulation of four-dimensional asymptotically locally anti-de Sitter and asymptotically locally flat spacetimes. 

Fefferman and Graham proposed an alternative gauge in \cite{PMP-FG1, PMP-FG2}, suitable for asymptotically anti-de Sitter spacetimes, but singular in the Ricci-flat limit. In this gauge the radial coordinate is space-like. It defines a family of time-like hypersurfaces, which asymptotes to the conformal boundary. The role of the Fefferman--Graham gauge in holography was recognized in the early developments of this field as a tool for unravelling the conformal class of boundary metrics together with the boundary conformal energy--momentum tensor \cite{Balasubramanian:1999re, Skenderis:2000in}.\footnote{Extend the Fefferman--Graham gauge so that Weyl covariance be manifest has been achieved only recently in Ref.  \cite{Ciambelli:2019bzz}.} 

Although, contrary to Fefferman--Graham, the Bondi gauge has not been significant in holography, it has common features with the derivative expansion of fluid/gravity correspondence: both are of the Eddington--Finkelstein type with one null radial coordinate and a retarded time \cite{Bhattacharyya:2007, Haack:2008cp, Bhattacharyya:2008jc, Hubeny:2011hd, Caldarelli:2012cm, Mukhopadhyay:2013gja, Petropoulos:2014yaa, Gath:2015nxa, Petropoulos:2015fba}. This intimate relationship holds in the conventional AdS/CFT holography as well as in the more recent and embryonic Ricci-flat/Carrollian-field-theory limit of the former, in its Ricci-flat-gravity/Carrollian-fluid emanation \cite{CMPPS2, CCMPS}.  

This latter viewpoint arouses new challenges around the Bondi gauge. Not only should we further delve into its rather novel anti-de Sitter side and understand the mass and angular momentum aspects, the news tensor, the asymptotic symmetries etc., but also translate the properties of the bulk in terms of the boundary geometric and hydrodynamical data, irrespective of the situation -- asymptotically AdS or flat. This last feature is utterly unexplored, and three dimensions provide again a  vast arena.  

The motivations of the present work on three-dimensional Einstein gravity are multiple, and concern evenly locally anti-de Sitter and locally flat spacetimes. At the first place we would like to discuss the complete solution spaces, as they appear from fluid/gravity correspondence, in Bondi, or in Fefferman--Graham gauge (when applicable). Our aim for such an exhaustive analysis is bound to the fact that the solution spaces are the antechamber for determining the asymptotic charges and their
general algebras. 
According to \cite{Troessaert:2013fma, Perez:2016vqo, Grumiller:2016pqb, Grumiller:2017sjh,oblak}, these algebras are expected to be bigger than the standard double Virasoro or $\mathfrak{bms}_3$, 
but concrete realizations in terms of solutions are rather sparse.\footnote{In Ref. \cite{CCMPS}, specific corners of the solution space were illustrated within fluid/gravity correspondence. These exhibit indeed different subalgebras of the expected complete algebra of asymptotic charges.} Prior to investigating the algebras, we need to unveil the residual diffeomorphisms and this is our second task, which goes along with setting the precise diffeomorphisms required to pass from one gauge to another.  This last step enables us to clarify the interplay between the Bondi gauge and the derivative expansion of fluid/gravity, and describe Bondi data in terms of boundary fluid variables, which is our third intent.  

The general solution space emerging from fluid/gravity correspondence is analyzed in Sec. \ref{FGDE}, grounded in two-dimensional hydrodynamics, which we recall for that purpose. This study covers both relativistic and Carrollian fluids, together with their gravity-dual anti-de Sitter or Minkowski spacetimes. Six arbitrary functions of two boundary coordinates define the solution spaces, and the residual diffeomorphisms are generated by four functions, for which we provide the variations and composition rules.
The Bondi gauge is introduced in Sec. \ref{BG}, accompanied with its solution space (five functions) and the corresponding residual diffeomorphisms (three functions). Its relation to the fluid/gravity gauge is also discussed there, and amounts to simply switching off one of the six functions present in the fluid/gravity description. This function is one component of the fluid velocity, and consequently the Bondi gauge amounts to choosing a specific hydrodynamic frame. The coordinate transformation necessary to reach the Bondi gauge from any point of the fluid/gravity solution space is a specific residual diffeomorphism of the latter, generating a change of hydrodynamic frame, i.e. a local Lorentz or Carrollian boost. We exhibit explicitly this diffeomorphism. Finally, Sec. \ref{FGG} is devoted to the Fefferman--Graham gauge, following the usual pattern: solution space (five functions), residual diffeomorphisms (three functions) and the explicit coordinate transformation necessary to reach Fefferman--Graham from Bondi. An appendix supplements our exposition with some detailed expressions for the Bondi gauge (App. \ref{app1}), and a note (App. \ref{app2}) on the algebra of residual diffeomorphisms, as it emerges from our study in Sec.  \ref{genbulk}, nicely fitting the results available in the current literature. An alternative and complementary presentation, including a useful Mathematica notebook and summarizing the current analysis is also available in the conference contribution  \cite{CMPRpos}.
 
\section{The fluid/gravity correspondence and its derivative expansion}
\label{FGDE}

\subsection{From the boundary \dots} \label{FGDEbdy}

\subsubsection*{The standard relativistic fluids}

A relativistic fluid flows on a pseudo-Riemannian spacetime along a  congruence $\text{u}$ with norm $\| \text{u} \|^2=-k^2$ ($k$ plays here the role of velocity of light and will be related in the next section to the bulk cosmological constant). The heat current $\text{q}$ being transverse, it is aligned in two dimensions with the Hodge-dual\footnote{Our conventions are:  $\ast u_\rho=u^\sigma \eta_{\sigma\rho}$ with $ \eta_{\sigma\rho}=\sqrt{\vert\det g\vert} \epsilon_{\sigma\rho}$ and $\epsilon_{01}=+1$ (there is a minus sign with respect to \cite{CMPRpos}). Hence $\eta^{\mu\sigma}\eta_{\sigma\nu}=\delta^\mu_\nu$.  The Hodge-dual of a vector is the Hodge-dual of the form with the index raised. Vectors are spelled with plain text letters ($\text{u}$), while one-forms will be displayed in ordinary boldface ($\mathbf{u}$). 
}  $\ast\text{u}$, normalized as $\| \ast\text{u} \|^2=k^2$:
\begin{equation}
\label{chimag}
\text{q}=\chi \ast \text{u} \quad\text{with} \quad \chi=-\frac{1}{k^2}\ast u^\mu T_{\mu\nu}u^\nu
\end{equation}
the local \emph{heat density}, appearing here as the magnetic dual of the \emph{energy density}
\begin{equation}
\label{long} 
\varepsilon=\frac{1}{k^2}T_{\mu \nu} u^\mu u^\nu.
\end{equation}
In these expressions, we have used the energy--momentum tensor $\text{T}=T_{\mu \nu}\text{d}x^\mu\text{d}x^\nu$, which is symmetric. The spacetime metric $\text{d}s^2=g_{\mu\nu}\text{d}x^\mu\text{d}x^\nu$ can also be expressed in the Cartan coframe $\{\mathbf{u}, \ast \mathbf{u} \}$ as
\begin{equation}
\text{d}s^2
=
\frac{1}{k^2}\left(-\mathbf{u}^2+
 \ast \mathbf{u}^2
\right).
\label{ds2gen}
\end{equation}
The energy--momentum tensor takes the form:
\begin{equation}
\text{T}=\frac{1}{2k^2}\left(\left(\varepsilon+\chi\right)\left(\mathbf{u}+\ast \mathbf{u}\right)^2+\left(\varepsilon-\chi\right)\left(\mathbf{u}-\ast \mathbf{u}\right)^2\right)
+\frac{1}{k^2}(p-\varepsilon+\tau) \ast \mathbf{u}^2,
\label{Tgen}
\end{equation}
where, $\tau$ is the \emph{viscous stress scalar}, unique component of the viscous stress tensor 
\begin{equation}\label{viscstr}
 \tau_{\mu \nu}= \tau  h_{\mu \nu} \quad\text{with} \quad h_{\mu\nu}=\dfrac{1}{k^2}\ast u_{\mu} \ast u_{\nu}
\end{equation}
the projector onto the space transverse to the velocity field. The trace reads: 
$T^\mu_{\hphantom{\mu}\mu}=p-\varepsilon+\tau$.

It is admitted -- and extensively discussed in the literature (see e.g. \cite{Landau, RZ, Romatschke:2009im, Kovtun:2012rj, Ciambelli:2017wou, Grozdanov:2019kge, Kovtun:2019hdm}) -- that one can perform local Lorentz boosts on the velocity congruence, while keeping intact the energy--momentum tensor and the entropy current. This freedom may be reduced by setting some constraint, locking the fluid in a specific \emph{hydrodynamic frame}. At this stage, we wish to keep the freedom on the fluid velocity field complete: on the one hand, because a hydrodynamic-frame transformation is not totally innocuous (see \cite{Ciambelli:2017wou}), in particular regarding global properties, as shown in  \cite{CCMPS}; on the other hand, because the core of the present work relies on the control of the bulk gauge freedom, and a boundary velocity transformation amounts to a specific bulk diffeomorphism, which we will exhibit in the next section.  

There is no shear or vorticity in two spacetime dimensions. 
The only non-vanishing first-derivative  tensors of the velocity are the expansion scalars
$\Theta=\nabla_\mu  u^\mu
$  
and $\Theta^{\ast}=\nabla_\mu  \ast u^\mu 
$,
equivalently defined as the exterior derivatives of the velocity forms\footnote{Some remarks on notation are necessary in order to avoid confusion.
The hodge-dual of a scalar spells with a suffix star and is a two-form; $\Theta^{\ast}$ is just another scalar. For any vector $\text{v}$ and a function $h$, $\text{v}(h)$ stands for $v^\mu\partial_\mu h$. We remind the following identities: 
$\text{d}h= \frac{1}{k^2}\left(\ast \text{u}(h)\ast \mathbf{u}- \text{u}(h)\mathbf{u} \right)$,
$\ast\text{d}h= \frac{1}{k^2}\left(\ast \text{u}(h)\mathbf{u}-\text{u}(h) \ast  \mathbf{u} \right)$, $\text{d}^\dag\mathbf{w}=\ast\text{d}\ast\mathbf{w}=-\nabla^\mu w_\mu$ and
$\Box h=-\text{d}^\dag\text{d}h= \frac{1}{k^2}\left(\ast \text{u}(\ast \text{u}(h))+\Theta^\ast \ast \text{u}(h)-\text{u}(\text{u}(h))-\Theta \text{u}(h) \right)$. We quote also $\ast\left(\mathbf{u}\wedge \ast\mathbf{u}\right)=k^2$.} 
\begin{equation}
\text{d}\ast\mathbf{u} =\frac{\Theta}{k^2}\ast \mathbf{u} \wedge\mathbf{u} 
\quad \text{and} \quad
\text{d}\mathbf{u} =\frac{\Theta^\ast}{k^2}\ast \mathbf{u} \wedge \mathbf{u},
 \label{def21eq}
\end{equation}
or in the Lie bracket of the velocity vectors
\begin{equation}
\left[\text{u},\ast\text{u} 
\right]=
\Theta^\ast 
\text{u} 
-\Theta
\ast\text{u}.
 \label{def21eqv}
\end{equation}

All information about the Levi--Civita connection in the frame $\{\text{u}, \ast \text{u} \}$ is encapsulated in $\Theta$ and $\Theta^\ast$. In particular, the acceleration is expressed as
$
a_\mu =u^\nu \nabla_\nu u_\mu  =\Theta^{\ast}
\ast u_\mu
$. These data  can be combined into the Weyl connection one-form 
\begin{equation}
\label{Wconc}
\mathbf{A}=\frac{1}{k^2}\left(\mathbf{a} -\Theta \mathbf{u}\right)=
\frac{1}{k^2} \left( \Theta^{\ast} \ast\mathbf{u}  -\Theta \mathbf{u}  \right).
\end{equation}
The corresponding field strength is a two-form with a Hodge-dual scalar
\begin{equation}
\label{scalF}
F=\ast\text{d}\mathbf{A}=\eta^{\mu\nu}\partial_\mu A_\nu=\frac{1}{k^2}\left(\ast\text{u}(\Theta)
-\text{u}(\Theta^\ast) \right).
\end{equation}
This scalar features the Weyl curvature of the two-dimensional geometry. The ordinary Christoffel--Riemann curvature is    
\begin{equation}
\label{Wcurv}
R=2\text{d}^\dagger \mathbf{A}=
\frac{2}{k^2}\left(\text{u}(\Theta)
+\Theta^2
 - \ast \text{u}(\Theta^\ast) 
- \Theta^{\ast 2}\right)
.
\end{equation}

Under Weyl transformations $\text{d}s^2\to  \nicefrac{\text{d}s^2}{{\cal B}^2}$
the velocity-form components $u_\mu$ are mapped to $\nicefrac{u_\mu}{{\cal B}}$. The Weyl connection one-form transforms as $\mathbf{A}\to\mathbf{A}-\text{d}\ln {\cal B}$, and its scalar field strength $F$ has weight 2 -- as opposed to the Christoffel--Riemann scalar, which has a weight-2 \emph{anomalous} transformation. In order to preserve the Weyl transformation properties of a conformal tensor,  
the ordinary covariant derivative $\nabla$   
should be traded for the Weyl-covariant combination $\mathscr{D}=\nabla+w\,\mathbf{A}$, $w$ being the conformal weight of the tensor. 

The dynamics of a relativistic fluid, subject to an external force of density $f_\nu$ is captured by the equations 
\begin{equation}
\label{T-cons}
\nabla^\mu T_{\mu\nu}=f_\nu,
    \end{equation}
supplemented with an equation of state. In holographic systems, the local-equilibrium equation of state is conformal,
$\varepsilon=p$, and the energy as well as heat densities have weight 2.  
For non-vanishing viscous stress scalar, the fluid is conformal  only when global equilibrium is assumed.

Anticipating the analysis of Sec. \ref{genbulk}, we will here introduce a special class of fluids, dubbed holographic, for which 
\begin{equation}
\label{anomaly}
\tau=\frac{R}{8\pi G}=
\frac{1}{4\pi Gk^2}\left(\text{u}(\Theta) +
\Theta^2- \ast \text{u}(\Theta^\ast) - \Theta^{\ast 2}
\right)
\end{equation}
and
\begin{equation}
\label{force-b}
f_\nu=-\nabla^\mu D_{\mu\nu}
\end{equation}
with $D_{\mu\nu}$ the components of the following symmetric and traceless tensor: 
\begin{equation}
\label{D}
D_{\mu\nu}\text{d}x^\mu \text{d}x^\nu =\frac{1}{8\pi G k^4}
\left(\left(\text{u}(\Theta) + \ast \text{u}(\Theta^\ast)-\frac{k^2}{2}R\right) 
\left(\mathbf{u}^2+
 \ast \mathbf{u}^2
\right)-4\ast\text{u}(\Theta)\mathbf{u} \ast \mathbf{u}
\right).
\end{equation}
The force vanishes if and only if the boundary geometry is flat and Weyl-flat.
Combining \eqref{T-cons}, \eqref{anomaly} , \eqref{force-b} and \eqref{D}, we find the longitudinal (energy) and transverse (momentum) fluid equations: 
\begin{equation}
\label{T-cons-el-mag-nc-force} 
 \begin{cases}
\mathcal{L}\equiv 
\text{u}(\varepsilon)+ 2\Theta \varepsilon
+ \ast\text{u}(\chi)+ 2\Theta^\ast\chi
-\frac{1}{4\pi G} \left[\ast\text{u}(F)+ 2\Theta^\ast F\right]=0,
\\
k\mathcal{T}\equiv\ast\text{u}(\varepsilon)+ 2\Theta^\ast \varepsilon
+
\text{u}(\chi)+ 2\Theta \chi
 =
0.
\end{cases}
\end{equation} 
These equations are Weyl-covariant of weight $3$ . 

We conclude this rapid overview of the two-dimensional relativistic dynamics with a generic parameterization of $\mathbf{u}$ and $\ast\mathbf{u}$, in terms of four arbitrary functions, $\Gamma$, $\Delta$, $v^\phi$ and $\gamma$, of two coordinates $\{u,\phi\}$. This will be useful, when dealing with a specific gauge, as in Sec. \ref{BG}. The expressions for the forms are\footnote{Using Eqs. \eqref{def21eq} we obtain the relativistic expansions: $
\Theta^\ast=\frac{k}{\Gamma}\left(-\partial_\phi\ln\gamma+\gamma\left(\partial_u \Delta+\partial_\phi\left(\Delta v^\phi\right)\right)\right)
$
and
$
\Theta=\frac{\gamma}{\Gamma}\left(\partial_u\Gamma
+\partial_\phi\left(\Gamma v^\phi\right) 
\right)
$.\label{relexp}
} 
\begin{equation}
\label{genf}
\mathbf{u} = k^2 \left(-\frac{\text{d}u}{\gamma}+\Delta\left(\text{d}\phi-v^\phi\text{d}u\right)\right),
\quad \ast \mathbf{u} = k\Gamma \left(\text{d}\phi-v^\phi\text{d}u\right),
\end{equation}
and equivalently for the vectors
\begin{equation}
\label{genv}
\text{u} =\gamma\left(\partial_u+v^\phi\partial_\phi\right),
\quad \ast \text{u} = \frac{k}{\Gamma} \left(\partial_\phi
+\Delta \gamma \left(\partial_u+v^\phi\partial_\phi\right)
\right).
\end{equation}
Among the four functions, $v^\phi$ and $\gamma$ have a clear physical meaning:  the physical velocity of the fluid and its Lorentz factor.
The boundary metric \eqref{ds2gen} reads:
\begin{equation}
\label{genmetr}
\text{d}s^2
=-k^2 \frac{\text{d}u^2}{\gamma^2}
+2k^2 \frac{\Delta}{\gamma}\text{d}u\left(\text{d}\phi-v^\phi\text{d}u\right)
+\left(\Gamma^2-k^2 \Delta^2\right)
\left(\text{d}\phi-v^\phi\text{d}u\right)^2.
\end{equation}

\subsubsection*{The more exotic Carrollian fluids}

The Carrollian world emerged with the seminal work of  L\'evy--Leblond \cite{Levy}. Although kinematically restricted due to the vanishing velocity of light (here $k$), the corresponding symmetry is as big as for Galilean systems, and provides a rich palette of mathematical  \cite{Henneaux:1979vn, Duval:1990hj, Barnich:2009se, Duval:2014uoa, Duval:2014uva, Duval:2014lpa, Ashtekar:2014zsa, Bekaert:2014bwa, Bekaert:2015xua, Hartong:2015xda, Figueroa-OFarrill:2018ilb, Morand:2018tke,Ciambelli:2019lap} and physical \cite{Bergshoeff:2014jla, card, ba, Penna1, CM1, Donnay:2019jiz} applications, mostly in relation with asymptotic symmetries of Ricci-flat gravitational backgrounds, and possibly with their holographic duals \cite{Bagchi2010e, Bagchi2, Att1, Att2,   Ball:2019atb, Merbis:2019wgk}. Assuming the latter exist, the study of their hydrodynamic regime calls for a theory of Carrollian fluids. 
Discussions or attempts for designing Carrollian (or generalized Galilean) hydrodynamics can be found in \cite{Hassaine:1999hn, Horvathy:2009kz, Penna2, Penna3, dutch, Poovuttikul:2019ckt, Armas:2019gnb, dutch2}. A comprehensive study was performed in \cite{CMPPS1}. This is based on the systematic analysis of relativistic hydrodynamics in the limit of vanishing light velocity. It has led to a set of Carrollian fluid equations, proven successful when  applied to the description of flat holography \cite{CMPPS2, CCMPS}. 
Going beyond the hydrodynamic regime requires a deeper understanding of Carrollian field theory, yet at a primitive stage (see e.g.  \cite{Bagchi:2019xfx}).

Carrollian fluids ``flow'' on Carrollian manifolds. The intrinsic definition of a Carrollian spacetime is very precise: it is a fiber bundle equipped with a degenerate metric of one-dimensional kernel and an Ehresmann connection. What a Carrollian fluid is, from first principles, is  not so clear because motion is forbidden when the light cone shrinks. For this reason we rather consider it as a limiting instance of a relativistic fluid. The zero-$k$ limit is not a priori well defined, and assumptions should be made both for the fluid and for the metric of the original spacetime. 

It is possible to choose a specific gauge (like the Papapetrou--Randers of \cite{CMPPS1}, or the light-cone coordinates introduced in  \cite{CCMPS}) so that, in the Carrollian limit, the fluid equations are manifestly covariant under \emph{Carrollian diffeomorphisms} i.e. 
coordinate transformations such that  $u'=u'(u, x)$ and $x^{\prime}=x^{\prime}(x)$. In these coordinates $u$ is the time, and the vector $\partial_u$ defines the kernel of the Carrollian metric.\footnote{Usually time is spelled $t$. However, when the holographic map is realized in Eddington--Finkelstein coordinates, the boundary time is associated with the bulk advanced time $u$, which should not be confused with the boundary fluid velocity vector $\text{u}$ or form $\mathbf{u}$, the components of the latter being $u^u$, $u^\phi$ and $u_u$, $u_\phi$.} In the systems at hand, space has a single direction, carried by $\partial_\phi$ (note that $\phi$ is not necessarily compact).
We will nevertheless refrain from locking the gauge at this stage. The equations will be less explicit but more convenient for our purposes.  

The kinematics of the fluid goes along with the geometry (it is encoded in the metric -- see  \eqref{ds2gen}). We must therefore assume the appropriate behavior of the forms $\{\mathbf{u},\ast \mathbf{u} \}$ and vectors $\{\text{u},\ast \text{u}\}$ for the Carrollian limit to exist, and use this behavior to define 
their Carrollian counterparts. This is inferred from the simplest example of a fluid at rest in flat spacetime with metric $\text{d}s^2=-k^2\text{d}u^2+\text{d}\phi^2$; hence the forms read $\mathbf{u}=-k^2 \text{d}u$ and $\ast \mathbf{u}=k \text{d}\phi$, whereas the corresponding vectors are  $\text{u}=\partial_u $ and $\ast \text{u}=k \partial_\phi$. With this, we define the Carrollian forms
\begin{equation}
\label{limudc}
\pmb{\mu}=\lim_{k\to 0}\frac{\mathbf{u}}{k^2}
 ,
\quad
\ast\pmb{\mu}=\lim_{k\to 0}\frac{\ast\mathbf{u}}{k},
\end{equation}
so that\footnote{Slanted bold fonts will be generally used to designate Carrollian forms.
They will carry different names since no map exists between tangent and cotangent spaces -- the metric is degenerate.}
\begin{equation}
\label{carmet}
\text{d}\ell^2=\lim_{k\to 0}\text{d}s^2=\ast\pmb{\mu}^2
\end{equation}
is the Carrollian metric. Similarly, for the vectors, the prescription is 
\begin{equation}
\label{limuvecc}
\upsilon=\lim_{k\to 0}\text{u},
\quad
\ast\upsilon=\lim_{k\to 0}\frac{\ast\text{u}}{k}.
\end{equation}
These obey 
\begin{equation}
\pmb{\mu}(\upsilon)=-1,\quad
\ast\pmb{\mu}(\ast\upsilon)=1,\quad
\ast\pmb{\mu}(\upsilon)=0,\quad
\pmb{\mu}(\ast\upsilon)=0,\quad
\end{equation}
so that the Carrollian-time direction $\upsilon$ is the kernel of the degenerate metric \eqref{carmet} (i.e. the vertical subbundle of the Carrollian manifold) -- see also \cite{Hartong:2015xda, Hartong:2015usd}. Due to this degeneracy, forms and vectors are not related to each other by lowering/raising indices, and this is why we have assigned  different symbols to them. Simultaneously, we have implicitly defined the Hodge duality, that cannot be introduced with its usual definition because the determinant of the metric vanishes.\footnote{In order to make this definition complete, we should further require that $\pmb{\mu}=\ast\ast\pmb{\mu}$ as well as $\ast\left(\pmb{\mu}\wedge \ast \pmb{\mu}\right)=1$, and similarly for the vectors. We will not expand these formal issues here. We refer instead to the appendix A of \cite{Hartong:2015usd}, where the Carrollian Hodge dual at hand was defined -- Eq. (A.41).} 

Using the scalings \eqref{limudc},  \eqref{limuvecc}, as well as the definitions \eqref{def21eq},  \eqref{def21eqv}, we  reach the Carrollian expansions $\theta$ and $\theta^\ast$ via
\begin{equation}
\text{d}\ast\pmb{\mu} =\theta\ast \pmb{\mu} \wedge\pmb{\mu} 
\quad \text{and} \quad
\text{d}\pmb{\mu} =\theta^\ast \ast\pmb{\mu} \wedge \pmb{\mu},
 \label{def21eqc}
\end{equation}
or with the Lie bracket of the Carrollian velocity vectors
\begin{equation}
\left[\upsilon,\ast\upsilon 
\right]=
\theta^\ast 
\upsilon
-\theta
\ast\upsilon.
 \label{def21eqcv}
\end{equation}
These are related
to the relativistic expansions through\footnote{The scalar $\theta$ defined here is slightly more general than previously introduced in \cite{CMPPS2, CCMPS, CMPPS1}. It accounts for extra contributions, which were separately displayed in those references.} 
\begin{equation}
\label{limcarexplim}
\theta=\lim_{k\to 0}\Theta,
\quad
\theta^\ast = \lim_{k\to 0}\frac{\Theta^\ast}{k}.
\end{equation}
Furthermore, the Carrollian spacetime defined with the forms $\pmb{\mu}$, $\ast \pmb{\mu} $, vectors $\upsilon$, $\ast \upsilon$ and degenerate metric $\text{d}\ell^2$ is naturally equipped with a Carrollian Weyl connection and its descendent Carrollian curvature scalars. These are obtained using the vanishing-$k$ limit of the relativistic data, but everything could be 
defined from first Carrollian principles -- and will be ultimately expressed in terms of the above Carrollian building blocks. We obtain the Carrollian Weyl connection
\begin{equation}
\pmb{A}=\lim_{k\to 0}\mathbf{A}=
  \theta^{\ast} \ast\pmb{\mu}  -\theta\pmb{\mu}
\end{equation}
and its Carrollian curvature (the minus sign is conventional)
\begin{equation}
\label{skF}
s=-\lim_{k\to 0}kF=\upsilon(\theta^{\ast})-
\ast\upsilon(\theta)=-\ast\text{d}\pmb{A}.
\end{equation}
The ordinary Riemann--Christoffel curvature $R$ defined in \eqref{Wcurv} is singular, and allows to define two Carrollian curvature scalars as the coefficients of the terms of order $\nicefrac{1}{k^2}$ and 1 respectively: $
r_{\mathcal{L}}=
2\left(\upsilon(\theta)
+\theta^2
\right) $ and 
$
r_{\mathcal{T}}=2\left(
  \ast \upsilon(\theta^\ast) 
+ \theta^{\ast 2}\right)
$, where the indices stand for \emph{longitudinal} and \emph{transverse} with respect to $\upsilon$ (kernel of the Carrollian metric). These scalars can also be defined from first Carrollian principles as pieces of a curvature introduced in \cite{CMPPS1}, 
or alternatively as parts of a torsion  \cite{Hartong:2015usd}.

A Carrollian fluid has dynamical variables inherited from the ancestor relativistic hydrodynamics, namely energy and heat density. Following \cite{CCMPS,CMPPS1}, we will assume that in the zero-$k$ limit, the energy density is finite and non-vanishing. In order to avoid cluttering of notation, we will keep the same symbol $\varepsilon$ for the Carrollian variable. Regarding the heat current, it must vanish linearly so that
\begin{equation}
\label{chipi}
\zeta=\lim_{k\to 0} \frac{\chi}{k}
\end{equation}
will play the role of Carrollian heat density.\footnote{This quantity was spelled $\chi_\pi$ in \cite{CCMPS}, referring to a Carrollian heat current $\pmb{\pi}$ introduced in that work.} 

Given the above kinematical data $\pmb{\mu}$, $\ast\pmb{\mu}$, $\upsilon$, $\ast\upsilon$ and the dynamical variables $\varepsilon$, $\zeta$ describing a Carrollian fluid flowing over a Carrollian manifold, the holographic fluid equations are obtained as the Carrollian limit of the longitudinal $\mathcal{L}$ and transverse $\mathcal{T}$ relativistic equations\footnote{The factor $k$ in the definition of $\mathcal{T}$ (Eq. \eqref{T-cons-el-mag-nc-force}) is instrumental for delivering a sensible Carrollian limit in the momentum equation. A similar precaution is necessary for dealing with the Galilean limit (see the standard textbook  \cite{Landau}, or \cite{CMPPS1} for a simultaneous and general treatment of Galilean and Carrollian limits).} given in 
\eqref{T-cons-el-mag-nc-force}:
\begin{equation}
\label{T-cons-el-mag-nc-car} 
 \begin{cases}
\lim_{k\to 0} \mathcal{L}= 
\upsilon(\varepsilon)+2 \theta\varepsilon+\frac{1}{4\pi G}\left[
\ast\upsilon(s)+2 \theta^\ast  s
\right]
=0,
\\
\lim_{k\to 0} \mathcal{T}=
\ast\upsilon(\varepsilon)+2 \theta^\ast  \varepsilon+
\upsilon(\zeta)+2 \theta\zeta
=0.
\end{cases}
\end{equation} 
An alternative version of Carrollian fluid equations can be found in Refs. \cite{CCMPS,CMPPS1}. There,  the equations are displayed in the Papapetrou--Randers gauge, where the invariance under Carrollian diffeomorphisms and conformal transformations is manifest.

Following the parameterization \eqref{genf} and \eqref{genv} for the relativistic data, we  present here their Carrollian relatives:\footnote{The Carrollian expansions are defined in Eqs. \eqref{def21eqc}. They read: $
\theta^\ast=\frac{1}{\Gamma}\left(-\partial_\phi\ln\gamma+\gamma\left(\partial_u \Delta+\partial_\phi\left(\Delta v^\phi\right)\right)\right)
$
and
$
\theta=\frac{\gamma}{\Gamma}\left(\partial_u\Gamma
+\partial_\phi\left(\Gamma v^\phi\right) 
\right)
$.\label{carexp}
} 
\begin{equation}
\label{genfvc}
\begin{split}
\pmb{\mu} = -\frac{\text{d}u}{\gamma}+\Delta\left(\text{d}\phi-v^\phi\text{d}u\right),
\quad \ast \pmb{\mu} = \Gamma \left(\text{d}\phi-v^\phi\text{d}u\right),\\
\upsilon  =\gamma\left(\partial_u+v^\phi\partial_\phi\right),
\quad \ast \upsilon  = \frac{1}{\Gamma} \partial_\phi
+ \frac{\Delta \gamma}{\Gamma}\left(\partial_u+v^\phi\partial_\phi\right).
\end{split}
\end{equation}
Among the four arbitrary functions $\Gamma(u,\phi)$, $\Delta(u,\phi)$, $v^\phi(u,\phi)$ and $\gamma(u,\phi)$, only two pertain to the Carrollian metric \eqref{carmet}, which takes the form
 \begin{equation}
\label{genmetrc}
\text{d}\ell^2
=\Gamma^2
\left(\text{d}\phi-v^\phi\text{d}u\right)^2.
\end{equation}

\subsection{\dots to the bulk}\label{genbulk}

\subsubsection*{Locally anti-de Sitter spacetimes}

The fluid/gravity correspondence maps relativistic fluid configurations onto Einstein spacetimes. In this holographic duality, the fluid flows on the conformal boundary of the asymptotically (locally) anti-de Sitter bulk. The metric of the latter is obtained in the form of a derivative expansion \cite{Bhattacharyya:2007, Haack:2008cp, Bhattacharyya:2008jc, Hubeny:2011hd}, inspired from the fluid homonymous expansion. 
The fluid derivative expansion consists in expressing the heat current and the stress tensor as  expansions in increasing derivatives of the fluid velocity and temperature fields (and possibly chemical potentials -- see \cite{Landau, RZ, Romatschke:2009im, Kovtun:2012rj}). Similarly, the derivative expansion of the bulk metric is set up order by order in inverse powers of the holographic coordinate $r$, which is a null radial coordinate. 

The fluid/gravity correspondence is not bijective. Not every Einstein space is dual to a relativistic fluid -- a simple counting of degrees of freedom upholds this statement. This reservation is lapsed the moment we allow for non-hydrodynamic modes, which bring about extra contributions in the fluid and metric expansions, not captured in velocity derivatives. It is customary to keep calling this a ``fluid/gravity derivative expansion,'' even though it is, strictly speaking, neither fluid, nor derivative, and use it as a framework to describe any Einstein spacetime. This direction has been pursued in a series of works \cite{ Caldarelli:2012cm, Mukhopadhyay:2013gja, Petropoulos:2014yaa, Gath:2015nxa, Petropoulos:2015fba}, dealing at the same time with the possibility of resumming the expansion (see below).

Concretely, the fluid/gravity derivative expansion appears
in an Eddington--Finkelstein form without complete gauge fixing -- as opposed to Bondi or Fefferman--Graham (see Secs. \ref{BG} and \ref{FGG}). At each order enter the boundary tensors of appropriate conformal weight, ensuring the invariance of the bulk with respect to boundary Weyl transformations.  These tensors are usually, but not necessarily, derivatives of the velocity field. The radial-evolution Einstein equations fix this expansion, whereas the constraint equations translate into the boundary fluid dynamics. 

In some general classes, the derivative expansion can be resummed. In those instances, the heat current and the stress tensor are exactly determined by geometric tensors, hence expressed as finite-order derivatives of elementary fields. This is expected to hold in arbitrary dimensions,  and has been demonstrated in four-dimensional bulk, where the boundary Cotton tensor is the fundamental geometric object that provides the fluid data.  

Three dimensions are peculiar because most geometric and fluid tensors vanish (like the shear or the vorticity). As a consequence, only a few quantities, compatible with conformal invariance remain.\footnote{Reminder: $\mathbf{u}$, $\ast \mathbf{u}$, $r$, $\varepsilon$ and $\chi$ have weights $-1$, $-1$, 1, 2 and 2.} These include the heat current, which enters freely and as an independent function the derivative expansion. The latter terminates at finite order and we find:
\begin{equation}
\text{d}s^2_{\text{Einstein}} =
2\frac{\mathbf{u}}{k^2}\left(\text{d}r+ \frac{r}{k^2} \left( \Theta^{\ast} \ast\mathbf{u}  -\Theta \mathbf{u}  \right)\right)+r^2\text{d}s^2+ \frac{8\pi G}{k^4} \mathbf{u}\left(\varepsilon \mathbf{u}  +\chi \ast \mathbf{u} \right),
\label{papaefgenrescrec3d}
\end{equation}
$\varepsilon = p$ and $\chi$ being 
the energy and heat densities of the fluid, $\mathbf{u}$ and $\ast\mathbf{u}$ its velocity and dual velocity form fields, and $\text{d}s^2$ the boundary metric expressed as in \eqref{ds2gen}. When inserted in
Einstein's equations with $\Lambda = -k^2$, adopting the triad $\{\text{d}r, \mathbf{u},\ast \mathbf{u}\}$ as Cartan coframe, the metric \eqref{papaefgenrescrec3d} solves
\begin{itemize}

\item the radial components, $rr$, $\mathbf{u}r$ and $\ast\mathbf{u}r$;

\item the transverse components, $\mathbf{u}\mathbf{u}$, $\ast\mathbf{u}\ast\mathbf{u}$ and $\ast\mathbf{u}\mathbf{u}$,
provided the fluid energy--momentum tensor \eqref{Tgen} obeys
\begin{equation}
\label{T-tilde-cons}
\nabla^\mu \left(T_{\mu\nu}+D_{\mu\nu}\right)=0,
    \end{equation}
where $D_{\mu\nu}$ are given in \eqref{D}, and the viscous stress scalar $\tau$ carries the conformal anomaly \eqref{anomaly}. The covariant derivative in \eqref{T-tilde-cons} is associated with the Levi--Civita connection of the boundary metric $\text{d}s^2$,  and Eqs. \eqref{T-tilde-cons} are equivalently spelled in the form \eqref{T-cons-el-mag-nc-force}.

\end{itemize}

According to the method of Ref. \cite{Balasubramanian:1999re}, the holographic energy--momentum tensor of the bulk metric \eqref{papaefgenrescrec3d} turns out to be the sum  
\begin{equation}
\label{T-tilde-em}
\tilde T_{\mu\nu}= T_{\mu\nu}+D_{\mu\nu}.
    \end{equation}
Hence, an alternative for the 
holographic-fluid energy--momentum tensor could have been $\tilde T_{\mu\nu}$. Decomposed as in  \eqref{Tgen}, the latter would have led to different energy and heat densities than  $\varepsilon$ and $\chi$, namely
\begin{equation}
\label{tilvarepch}
\begin{split}
\tilde\varepsilon&=\varepsilon
+\frac{1}{8\pi Gk^2}\left(\text{u}(\Theta) + \ast \text{u}(\Theta^\ast)\right) 
- \frac{R}{16\pi G},
\\
\tilde\chi&= \chi -\frac{1}{4\pi Gk^2}\ast\text{u}(\Theta)
,
\end{split}
\end{equation}
whereas $\tilde \tau=\tau$ because $D_{\mu\nu}$ has vanishing trace. This option would have rendered the expression for the bulk metric less natural, and somehow blurred its Carrollian limit (discussed in the next paragraph), because of divergences at vanishing $k$, occurring in the tilde energy and heat densities  (see \cite{CCMPS} for details). It is nonetheless important for discussing the boundary local Lorentz transformations i.e. the changes of holographic-fluid boundary frame, and their translation into bulk  diffeomorphisms.

A comment is worth making at this stage.  Even though all relevant information carried by the energy--momentum tensor  ($\varepsilon$, $\chi$, $\tau$, $\mathbf{u}$ and $\ast\mathbf{u}$) is used for the reconstruction of the bulk metric \eqref{papaefgenrescrec3d}, in the gauge at hand, the energy--momentum tensor does not appear in the metric as a single piece of holographic boundary data. This is to be contrasted to what happens in the 
Fefferman--Graham gauge, where the entire energy--momentum emerges at a specific order in the radial, holographic  expansion -- see Sec. \ref{FGG}.

Expression \eqref{papaefgenrescrec3d} is partly on-shell.  It depends on six functions: $\varepsilon(u,\phi)$, $\chi(u,\phi)$, the two components of $\mathbf{u}$ ($u_u(u,\phi)$ and $u_\phi(u,\phi)$) and those of $\ast\mathbf{u}$ ($\ast u_u(u,\phi)$ and $\ast u_\phi(u,\phi)$) -- as displayed e.g. in Eqs. \eqref{genf}. It exhibits the most general locally AdS spacetime in Eddington--Finkelstein coordinates, whenever these six functions obey \eqref{T-cons-el-mag-nc-force}. There is one more function than in Bondi gauge, as we will see in Sec. \ref{BG}, precisely because the derivative expansion is not constructed as a gauge, i.e. it ensures only a partial gauge fixing. In fact, the extra degree of freedom corresponds to the (local) arbitrariness of hydrodynamic frame, and is absent in Bondi gauge -- where the fluid flows in a frame dubbed \emph{Bondi hydrodynamic frame}.

Given \eqref{papaefgenrescrec3d}, it is legitimate to wonder what the residual diffeomorphisms are. These are transformations, which keep \eqref{papaefgenrescrec3d} form-invariant, while modifying its building blocks, $\mathbf{u}$, $\ast\mathbf{u}$, $\varepsilon$ and $\chi$. Our motivation for such an analysis  is twofold. At the first place this will set up the stage for the comparison with Bondi or Fefferman--Graham gauges. In addition, the set of residual diffeomorphisms is a prerequisite for determining the asymptotic charges and their algebra. With the universal form of the derivative expansion at hand, it is expected to recover the general algebra advertised in \cite{Troessaert:2013fma,Grumiller:2016pqb}, extending thereby the partial results obtained in \cite{CCMPS}.

Since the derivative expansion is a partial gauge fixing, the residual diffeomorphisms encompass an arbitrary function of all bulk coordinates \cite{Ruzziconi:2019pzd}. These are generated by a bulk vector field, expressed  either in the natural bulk frame $\{\partial_r, \partial_\mu\}$, or in the Cartan frame $\{\partial_r, \text{u}, \ast\text{u}\}$ that can be adopted for the bulk:
\begin{equation}
\label{Killxigen}
\xi = \xi^r \partial_r +   \xi^\mu \partial_\mu=\xi^r  \partial_r - \frac{u_\mu\xi^\mu}{k^2}\text{u}+\frac{\ast u_\mu\xi^\mu}{k^2}\ast\text{u}, 
\end{equation}
where all three components depend on $r$ and $x^\mu$. They can be expressed as an $\nicefrac{1}{r}$ power series. Inspired by the analysis performed in Bondi gauge, we make here the ansatz to terminate this series at first order:
\begin{equation}
\label{kilads}
\xi^r =  r\xi^r_{(-1)} +\xi^r_{(0)}+\frac{1}{r} \xi^r_{(1)} ,\quad
\xi^\mu=\xi^\mu_{(0)}+\frac{1}{r} \xi^\mu_{(1)}, 
\end{equation}
where $\xi_{(0)} = \xi^\mu_{(0)}\partial_\mu$ and $\xi_{(1)} = \xi^\mu_{(1)}\partial_\mu$ are boundary vector fields, and $\xi^r_{(-1)} $, $\xi^r_{(0)} $ as well as $\xi^r_{(1)} $ boundary scalars.
The subsequent analysis is based on the Lie derivative of the bulk metric along $\xi$; it is ultimately recast in boundary language. 

The condition\footnote{In order to avoid any confusion, we will generically  spell the bulk metric as 
$$
\text{d}s^2_{\text{bulk}}=G_{MN}\text{d}x^M\text{d}x^M=
G_{rr}\text{d}r^2
+2G_{r\alpha}\text{d}r\text{d}x^\alpha
+G_{\mu\nu}\text{d}x^\mu\text{d}x^\nu,
$$
where $G_{MN}$ are functions of all coordinates.
} $\mathscr{L}_\xi G_{rr}=0$ enforces transversality for $\xi_{(1)}$ with respect to $\text{u}$:
\begin{equation}
\xi _{(1)}=  \frac{Z}{k}\ast \text{u}
\end{equation}
with $Z$ an arbitrary boundary scalar function. The conditions stemming out from $\mathscr{L}_\xi G_{r\mu}$ result in
\begin{equation}
\xi^r_{(-1)}=S
,\quad 
\xi^r_{(1)}=-\frac{4\pi G }{k} \chi Z,
\end{equation}
where $S$ is another arbitrary boundary function, while $\chi$ is the boundary fluid heat density. The treatment of $\mathscr{L}_\xi G_{\mu\nu}$ imposes 
\begin{equation}
\xi^r_{(0)}=-\nabla_\mu \xi^\mu_{(1)}=- \frac{1}{k}\left(\Theta^\ast Z
 +\ast \text{u}(Z)  \right),
\end{equation}
whereas the boundary vector $\xi_{(0)}$ remains unconstraint and expressed in terms of two arbitrary functions $f$ and $Y$:
\begin{equation}
\label{Kilbdy}
\xi _{(0)}=  f \text{u} +\frac{Y}{k}\ast \text{u}.
\end{equation}

Summarizing, the residual diffeomorphisms are linearly encoded in four arbitrary functions of the boundary coordinates: $f$, $Y$, $S$ and $Z$. The components of their generating vector fields are
\begin{equation}
\label{kiladsgen}
\xi= \left(r S- \frac{1}{k}\left(\Theta^\ast Z
 +\ast \text{u}(Z)  \right)-\frac{4\pi G }{kr} \chi Z\right) \partial_r + f \text{u} 
+\frac{1}{k}\left(Y+\frac{Z}{r}\right)\ast\text{u}.
\end{equation}
The effect of these diffeomorphisms on the bulk metric \eqref{papaefgenrescrec3d} is reflected entirely  in the variation they produce on the boundary data, which uniquely define the bulk solution space.
These data are the velocity field and its dual
form, for which\footnote{Our convention for the variation $\delta_\xi $ is the opposite of that used in \cite{CMPRpos}.}
\begin{equation}
\label{deltaudexpl} 
\delta_\xi u_\mu = -\left(S +\frac{\Theta^\ast}{k}Y+\text{u}( f)
\right)u_\mu
+ k
\left(Z -\frac{\Theta^\ast}{k} f  +\frac{\ast\text{u}( f )}{k}
\right) \ast u_\mu
\end{equation}
and
\begin{equation}
\label{deltastudexpl} 
\delta_\xi \ast u_\mu = \frac{1}{k}
\left(k^2 Z -\Theta Y +\text{u}(Y)
\right)u_\mu
- 
\left(S +\Theta f  + \frac{\ast\text{u}(Y)}{k}
\right)\ast u_\mu,
\end{equation}
as well as the energy and heat densities:
\begin{equation}
\label{deltaeps}
\delta_\xi \varepsilon
=-Y\frac{\ast \text{u}(\varepsilon)}{k}- f 
 \text{u}(\varepsilon) +2 S\varepsilon-2kZ \chi+\frac{1}{4\pi G }\left[ kFZ -
  \Theta \frac{\ast \text{u}(Z)}{k}
 -\frac{ \text{u}(\ast \text{u}(Z))}{k}\right]
\end{equation}and 
\begin{equation}
\label{deltach}
\delta_\xi \chi
=-Y\frac{\ast \text{u}(\chi)}{k}- f 
 \text{u}(\chi) +2 S\chi-2kZ \varepsilon+\frac{1}{4\pi G }\left[\Theta^\ast \frac{\ast \text{u}(Z)}{k}
 + \frac{\ast \text{u}(\ast \text{u}(Z))}{k}\right].
\end{equation}
Under these transformations, the relativistic hydrodynamics equations \eqref{T-cons-el-mag-nc-force}, obeyed by the boundary data, remain unaltered.

The variations of the velocity fields \eqref{deltaudexpl} and \eqref{deltastudexpl} are of the generic form including a longitudinal and a transverse component
\begin{equation}
\delta_\xi u_\mu =- \psi u_\mu+\psi^\ast \ast u_\mu , \quad
\delta_\xi \ast u_\mu =\omega^\ast  u_\mu - \omega \ast u_\mu,
\label{deltuastud} 
\end{equation}
where $ \psi^\ast $, $ \psi $, $\omega$ and $\omega^\ast$ are read off directly in Eqs.  \eqref{deltaudexpl} and \eqref{deltastudexpl} (see also \eqref{Psist}, \eqref{Psi}, \eqref{W}, \eqref{Wst}). These four functions provide a choice for parameterizing a residual diffeomorphism, alternative to $f$, $Y$, $S$ and $Z$. The variations of the corresponding vectors and of the boundary metric read simply\footnote{We also quote: 
$\delta_\xi \Theta=
\psi \Theta+\omega^\ast \Theta^\ast+\ast\text{u}(\omega^\ast)-\text{u}(\omega)=
S \Theta -\text{u}(S)-f \text{u}(\Theta)-Y\frac{\ast \text{u}(\Theta)}{k}+k\left(Z\Theta^\ast+\ast \text{u}(Z)\right)
$ and $\delta_\xi \Theta^\ast=
\psi^\ast \Theta+\omega \Theta^\ast-\ast\text{u}\left(\psi\right))+\text{u}\left(\psi^\ast\right)=
S \Theta^\ast -\ast\text{u}(S)-f \text{u}(\Theta^\ast)-Y\frac{\ast \text{u}(\Theta^\ast)}{k}+k\left(Z\Theta+ \text{u}(Z)\right)$.\label{deltatheta}}
\begin{equation}
\delta_\xi u^\mu =\psi u^\mu+\omega^\ast \ast u^\mu , \quad
\delta_\xi \ast u^\mu =\psi^\ast  u^\mu+ \omega  \ast u^\mu,
\label{deltuastuU} 
\end{equation}
and
\begin{equation}
\delta_\xi \text{d}s^2=\frac{2}{k^2}\left(\psi\mathbf{u}^2
-\omega\ast\mathbf{u}^2
+\left(\omega^\ast -\psi^\ast
\right)\mathbf{u} \ast\mathbf{u}\right).
\label{delatxids2}
\end{equation}
A boundary Weyl transformation is therefore induced with $\omega^\ast=\psi^\ast$ and $\omega=\psi$. Ultimately, the boundary metric is invariant if furthermore $\omega=\psi=0$; $\psi$ measures therefore the change of scale in the metric (see App. \ref{app2} for a more elaborate discussion). We will now interpret $\psi^\ast$.

A local Lorentz transformation is a one-parameter subset of the residual diffeomorphisms that leave the boundary metric invariant. It acts on the velocity fields as
\begin{equation}
\label{locLor}
\begin{pmatrix}
\delta_{\text{L}} \text{u}\\
\delta_{\text{L}} \ast \text{u}\
\end{pmatrix}
=
 \psi^\ast\begin{pmatrix}
\ast \text{u}\\
\text{u}
\end{pmatrix}
\quad \text{and}
\quad
\begin{pmatrix}
\delta_{\text{L}} \mathbf{u}\\
\delta_{\text{L}} \ast \mathbf{u}\
\end{pmatrix}
=
 \psi^\ast\begin{pmatrix}
\ast \mathbf{u}\\
\mathbf{u}
\end{pmatrix}
,
\end{equation}
where anticipating the output, we have used the parameter $\psi^\ast$ appearing in Eqs. \eqref{deltuastud}, \eqref{deltuastuU}
and \eqref{delatxids2}.
This transfromation produces
\begin{equation}
\label{locLorThs}
\delta_{\text{L}} \Theta= \ast\text{u}\left(\psi^\ast\right)+\Theta^\ast \psi^\ast,\quad 
\delta_{\text{L}} \Theta^\ast=\text{u}\left(\psi^\ast\right)+\Theta \psi^\ast.
\end{equation}
By definition, this is the hydrodynamic-frame transformation, which keeps invariant the boundary geometry (i.e. the metric \eqref{ds2gen} and its Riemann--Christoffel curvature $R$ \eqref{Wcurv}), together with the energy--momentum tensor $\tilde T_{\mu\nu}$ given in \eqref{T-tilde-em}. The latter requirement sets (see \cite{CCMPS})
\begin{equation}
\label{locLor-en-he-nc}
\begin{pmatrix}
\delta_{\text{L}}\tilde\varepsilon\\
\delta_{\text{L}}\tilde\chi
\end{pmatrix}
=- 2\psi^\ast
\begin{pmatrix}
\tilde\chi\\
\tilde\varepsilon
\end{pmatrix}
- \psi^\ast
\begin{pmatrix}
 0\\
\tilde\tau
\end{pmatrix}, 
\end{equation}
while $\delta_{\text{L}}\tilde\tau=0$ because $\tilde\tau=\tau=\nicefrac{R}{8\pi G}$ (see \eqref{anomaly}). Applied to \eqref{tilvarepch}, the transformation rules \eqref{locLor}, \eqref{locLorThs} and \eqref{locLor-en-he-nc} lead to the actual energy and heat-density variations
\begin{equation}
\label{deltaepsL}
\delta_{\text{L}} \varepsilon
=-2\psi^\ast \chi+\frac{1}{4\pi G }\left[ F \psi^\ast  -
  \Theta \frac{\ast \text{u}\left(\psi^\ast\right)}{k^2}
 -\frac{ \text{u}(\ast \text{u}\left(\psi^\ast\right))}{k^2}\right]
\end{equation}and 
\begin{equation}
\label{deltachL}
\delta_{\text{L}} \chi
=-2\psi^\ast \varepsilon+\frac{1}{4\pi G }\left[\Theta^\ast \frac{\ast \text{u}\left(\psi^\ast\right)}{k^2}
 + \frac{\ast \text{u}(\ast \text{u}\left(\psi^\ast\right))}{k^2}\right].
\end{equation}
Comparing \eqref{locLor}, \eqref{deltaepsL}, \eqref{deltachL} with the general expressions
\eqref{deltuastud}, \eqref{deltaeps}, \eqref{deltach}, we conclude that
the hydrodynamic-frame transformations of the boundary fluid correspond to a subset of the bulk residual diffeomorphisms generated by a vector field \eqref{kiladsgen} with $f=Y=S=0$ and $\psi^\ast=kZ$:
\begin{equation}
\label{kiladsgenL}
\xi_{\text{L}}= - \frac{1}{k^2}\left(\Theta^\ast \psi^\ast
 +\ast \text{u}\left(\psi^\ast\right)\right)\partial_r  -\frac{\psi^\ast}{r} \left(\frac{4\pi G}{k^2}\chi \partial_r 
-\frac{\ast\text{u}}{k^2}\right).
\end{equation}
In the parameterization \eqref{genf}, the Lorentz transformation \eqref{locLor} acts as
\begin{equation}
\label{locLor-genf}
\delta_{\text{L}}\Gamma=k\psi^\ast \Delta,\quad
\delta_{\text{L}}\Delta=\frac{\psi^\ast\Gamma}{k},\quad
\delta_{\text{L}}v^\phi=\frac{k\psi^\ast}{\gamma\Gamma},\quad
\delta_{\text{L}}\gamma=\frac{k\psi^\ast\gamma\Delta}{\Gamma}.
\end{equation}

Considering again the complete family of residual diffeomorphisms parameterized by four functions $f$, $Y$, $S$ and $Z$, one may wonder how these are composed. The answer to this question requires the use of a modified Lie bracket  (see e.g.  \cite{Barnich:2010eb}), which suitably accounts for the effect that the geometry variation produces on the generators:
\begin{equation}
\label{modlie}
\xi_3=\left[\xi_1,\xi_2\right]_{\text{M}}=\left[\xi_1,\xi_2\right]+\delta_{\xi_1}\xi_2-\delta_{\xi_2}\xi_1.
\end{equation}
This bracket endows the family of generators \eqref{kiladsgen} with the structure of a Lie algebra, which in turn provides the composition rules of $f_1$, $Y_1$, $S_1$, $Z_1$ and $f_2$, $Y_2$, $S_2$, $Z_2$: 
\begin{eqnarray}
f_3&=&S_1 f_2-S_2f_1+Z_1Y_2-Z_2Y_1+\frac{\Theta^\ast}{k}\left(Y_1f_2-Y_2f_1\right)+\delta_{\xi_1}f_2-\delta_{\xi_2}f_1,
\label{AdS-algebra-f}
\\
Y_3&=&S_1 Y_2-S_2Y_1+k^2\left(Z_1f_2-Z_2f_1\right)+\Theta\left(f_1Y_2-f_2Y_1\right)+\delta_{\xi_1}Y_2-\delta_{\xi_2}Y_1,
\label{AdS-algebra-Y}
\\
S_3&=&f_1 \text{u}\left(S_2\right)-f_2 \text{u}\left(S_1\right)+\frac{1}{k}\left(Y_1 \ast\text{u}\left(S_2\right)-Y_2 \ast\text{u}\left(S_1\right)\right)+\delta_{\xi_1}S_2-\delta_{\xi_2}S_1,
\label{AdS-algebra-S}
\\
Z_3&=&f_1 \text{u}\left(Z_2\right)-f_2 \text{u}\left(Z_1\right)+\frac{1}{k}\left(Y_1 \ast\text{u}\left(Z_2\right)-Y_2 \ast\text{u}\left(Z_1\right)\right)+\delta_{\xi_1}Z_2-\delta_{\xi_2}Z_1.
\label{AdS-algebra-Z}
\end{eqnarray}
In these expressions, the last two terms ($\delta_{\xi_1}f_2-\delta_{\xi_2}f_1$ etc.) vanish whenever $f_a$, $Y_a$, $S_a$ and $Z_a$ are field-independent diffeomorphism parameters. 
This happens e.g. for the Lorentz boosts discussed above (where furthermore $Z_3=0$ as a manifestation of their abelian nature), but needs not be so in general. We will not elaborate any longer on the above algebra, some further information is available in App. \ref{app2}. It is very general and allows to recover some previous results  \cite{CCMPS, Troessaert:2013fma,Perez:2016vqo, Grumiller:2016pqb,oblak, Barnich:2009se} aiming at extending the standard Witt (or BMS -- see next paragraph) residual-symmetry algebras emerging in three spacetime dimensions. This analysis is the starting point for the determination of the surface charges.

\subsubsection*{Locally Minkowskian spacetimes}

In the conventional fluid/gravity holographic correspondence, the relativistic fluid flows on the conformal boundary.  In the limit of vanishing cosmological constant, the conformal boundary is traded for null infinity, which is indeed a Carrollian spacetime. Simultaneously, the relativistic fluid is mapped onto its Carrollian limit, defined at null infinity. One therefore expects that, for well-behaved relativistic fluid configurations, i.e. such that the limits \eqref{limudc}, \eqref{limuvecc} hold, the Carrollian counterparts should provide the building blocks for reconstructing asymptotically flat spacetimes with a Carrollian derivative expansion. This was successfully analysed in Refs. \cite{CMPPS2, CCMPS}. 
Indeed, under the assumptions recalled here, expression 
\eqref{papaefgenrescrec3d} is regular at vanishing $k$ with the limit  
\begin{equation}
\text{d}s^2_{\text{flat}} =
2\pmb{\mu}\left(\text{d}r+r \left(\theta^\ast \ast\pmb{\mu}-\theta\pmb{\mu}\right)\right)
+r^2\text{d}\ell^2
+8\pi G \pmb{\mu} \left(\varepsilon\pmb{\mu}+\zeta \ast\pmb{\mu}\right). 
\label{papaefresricfgen}
\end{equation}
The boundary (more precisely null-infinity) data are the Carrollian velocity forms $\pmb{\mu}$ and $ \ast\pmb{\mu}$ (their expansions $\theta$ and $\theta^\ast$ are defined in \eqref{def21eqc} or \eqref{def21eqcv}), the energy density $\varepsilon$ and the Carrollian heat density $\zeta $ (see \eqref{chipi}). The Carrollian degenerate metric $\text{d}\ell^2$ is built on $ \ast\pmb{\mu}$ (see \eqref{carmet}), and the Carrollian geometry is completed with the data $\upsilon$ and $\ast \upsilon $, associated with the Carrollian fluid velocity.

The metric \eqref{papaefresricfgen}, abusively called flat derivative expansion, has a finite number of terms and it is not completely off-shell as it satisfies $R_{rr}=R_{\pmb{\mu}r}= R_{\ast\pmb{\mu}r}=0$ (we use the Cartan triad $\{\text{d}r,\pmb{\mu},\ast\pmb{\mu}\}$). However, the remaining components of the bulk Ricci tensor $R_{\pmb{\mu}\pmb{\mu}}$, $R_{\ast\pmb{\mu}\ast\pmb{\mu}}$ and $R_{\ast\pmb{\mu}\pmb{\mu}}$ vanish if and only if the Carrollian fluid equations \eqref{T-cons-el-mag-nc-car} are satisfied. 

The above is a remarkable result, which provides the general expression of locally flat spacetimes in Eddington--Finkelstein coordinates. Again, six functions of two boundary coordinates describe the dynamics: $\pmb{\mu}=\mu_u(u,\phi)\text{d}u+\mu_\phi(u,\phi)\text{d}\phi$ and $\ast\pmb{\mu}=\ast \mu_u(u,\phi)\text{d}u+\ast \mu_\phi(u,\phi)\text{d}\phi$ (possibly parameterized following \eqref{genfvc}), as well as $\varepsilon(u,\phi)$ and $\zeta(u,\phi)$. As in AdS, this is one more than in Bondi gauge (see Sec. \ref{BG}).

The residual diffeomorphisms of the locally flat metric \eqref{papaefresricfgen} are parameterized in terms of the same functions as its anti-de Sitter counterpart  \eqref{papaefgenrescrec3d}. These are $f$, $Y$, $S$ and $Z$, and depend on the two boundary coordinates. They appear at the first place in the variations of the holographic data $\pmb{\mu}$, $\ast \pmb{\mu}$, $\varepsilon$ and $\zeta$, which in turn transform the bulk metric in a form-invariant manner. 

The generating vector fields for the residual diffeomorphisms are of the form \eqref{Killxigen}, and can be explicitly obtained through an analysis of the metric Lie derivatives, similar to that performed in the anti-de Sitter case. The result coincides with the zero-$k$ limit of \eqref{kiladsgen}, which can be spelled in the bulk vector frame $\{\partial_r, \upsilon, \ast\upsilon\}$:
\begin{equation}
\label{kiladsgencar}
\xi= \left(r S-\theta^\ast Z-\ast \upsilon(Z)-\frac{4 \pi G}{r} \zeta  Z\right) \partial_r + f  \upsilon
+\left(Y+\frac{Z}{r}\right)\ast \upsilon.
\end{equation}
Again, these vectors depend on four arbitrary functions. 
Using \eqref{deltaudexpl} and \eqref{deltastudexpl}) together with \eqref{limudc}, \eqref{limuvecc} and \eqref{limcarexplim}, we find
\begin{equation}
\begin{split}
\label{deltaudexpllim} 
\delta_\xi \pmb{\mu} &= -\left(S +\theta^\ast Y +\upsilon(f )
\right)\pmb{\mu}
+
\left(Z -\theta^\ast f  +\ast\upsilon(f)
\right) \ast \pmb{\mu},\\
\delta_\xi \ast \pmb{\mu} &=
\left(\upsilon(Y)
-\theta Y\right)\pmb{\mu}
-
\left(S+\theta f+\ast \upsilon(Y)
\right)\ast \pmb{\mu},
\end{split}
\end{equation}
as well as
\begin{equation}
\begin{split}
\label{deltauUexpllim} 
\delta_\xi \upsilon &= 
\left(S +\theta^\ast Y +\upsilon(f )
\right)\upsilon
+ \left(\upsilon(Y)-\theta Y
\right)
 \ast \upsilon,\\
\delta_\xi \ast \upsilon &=
\left(Z -\theta^\ast f  +\ast\upsilon(f)
\right)
\upsilon
+
\left(S+\theta f+\ast \upsilon(Y)
\right)\ast \upsilon.
\end{split}
\end{equation}
These relations enable us to write the transformation of the Carrollian metric:\footnote{Following footnote \ref{deltatheta}, we find here: $\delta_\xi \theta=S \theta -\upsilon(S)-f \upsilon(\theta)-Y\ast \upsilon(\theta)$ and $\delta_\xi \theta^\ast=S \theta^\ast -\ast\upsilon(S)-f \upsilon(\theta^\ast)-Y\ast \upsilon(\theta^\ast)+Z\theta+ \upsilon(Z)
$.}
\begin{equation}
\label{dl2lim} 
\delta_\xi \text{d}\ell^2 = 2\ast \pmb{\mu}\delta_\xi \ast\pmb{\mu}=-2 \left(S+\theta f+\ast \upsilon(Y)
\right)\text{d}\ell^2 
+2\left(\upsilon(Y)-\theta Y
\right)\pmb{\mu}\ast \pmb{\mu}
.
\end{equation}
The transformations at hand keep the Carrollian metric degenerate. They induce a Weyl transformation under the condition $\upsilon(Y)=\theta Y$ (see App. \ref{app2} for further details). If furthermore $S+\theta f+\ast \upsilon(Y)=0$, the bulk diffeomorphisms do not affect the boundary Carrollian metric; they include local Carrollian boosts, as we will see in a short while. 

Finally, starting from \eqref{deltaeps} and \eqref{deltach}, we obtain the  Carrollian counterparts of the energy and heat density transformations:
\begin{equation}
\label{deltaepscar}
\delta_\xi \varepsilon
=-Y\ast \upsilon\left( \varepsilon\right)-f 
\upsilon(\varepsilon) +2 S\varepsilon
-\frac{1}{4\pi G }\left[s Z +\theta
  \ast \upsilon(Z)
 + \upsilon\left( \ast \upsilon(Z) \right)\right]
\end{equation}
with $s$ being the Carrollian Weyl curvature \eqref{skF}, and 
\begin{equation}
\label{deltachcar}
\delta_\xi \zeta
=-Y\ast\upsilon\left( \zeta\right)-f 
 \upsilon\left( \zeta\right) +2 S \zeta-2Z \varepsilon+\frac{1}{4\pi G }\left[\theta^\ast\ast\upsilon(Z)
 +\ast\upsilon\left(\ast\upsilon(Z)\right)\right].
\end{equation}
The transformations under consideration respect 
the Carrollian fluid equations \eqref{T-cons-el-mag-nc-car}.

Before ending this paragraph, we should discuss the fate of the boundary local Lorentz transformations, i.e. the hydrodynamic-frame freedom in the Carrollian limit. Although by essence this freedom is lost, requiring the scaling of the parameter $\psi^\ast$ as $k\alpha$ in the transformation rules \eqref{locLor}, enables us to recover a non-trivial Carrollian remnant as  
\begin{equation}
\label{locCar}
\delta_{\text{C}} \pmb{\mu}=\alpha \ast \pmb{\mu},\quad
\delta_{\text{C}} \ast \pmb{\mu}=0,\quad
\delta_{\text{C}} \upsilon=0
,\quad
\delta_{\text{C}} \ast \upsilon=\alpha \upsilon,
\end{equation}
resulting in
\begin{equation}
\label{locCarths}
\delta_{\text{C}} \theta=0,\quad 
\delta_{\text{C}} \theta^\ast=\upsilon(\alpha)+\theta \alpha,
\end{equation}
and keeping the Carrollian metric \eqref{carmet} invariant (see \eqref{dl2lim}). This is a local Carrollian boost. Thanks to the scaling $\psi^\ast=k\alpha$, Eqs. \eqref{deltaepsL} and  \eqref{deltachL} have a smooth vanishing-$k$ limit,  
\begin{equation}
\label{deltaepsC}
\delta_{\text{C}} \varepsilon
=-\frac{1}{4\pi G }\left[  s\alpha +
  \theta \ast \upsilon(\alpha)
 +\upsilon(\ast \upsilon(\alpha))\right]
\end{equation}and 
\begin{equation}
\label{deltachC}
\delta_{\text{C}} \zeta
=-2\alpha \varepsilon+\frac{1}{4\pi G }\left[\theta^\ast \ast \upsilon(\alpha)
 + \ast \upsilon(\ast \upsilon(\alpha))\right].
\end{equation}
We can compare \eqref{locCar}, \eqref{deltaepsC} and \eqref{deltachC} with 
\eqref{deltaudexpllim}, 
\eqref{deltauUexpllim}, 
  \eqref{deltaepscar} and \eqref{deltachcar}. We observe that
the Carrollian descendants of the hydrodynamic-frame transformations are a subset of the bulk residual diffeomorphisms generated by \eqref{kiladsgencar} with $f=Y=S=0$ and $Z=\alpha$:
\begin{equation}
\label{kilmingenC}
\xi_{\text{C}}= - \left(\theta^\ast \alpha
 +\ast \upsilon(\alpha)\right)\partial_r  -\frac{\alpha}{r} \left(4\pi G\zeta \partial_r 
-\ast \upsilon\right).
\end{equation}
This is indeed the Carrollian limit of \eqref{kiladsgenL}. Again, using the parameterization \eqref{genfvc}, the transformation \eqref{locCar} results in
\begin{equation}
\label{locCar-genfvc}
\delta_{\text{C}}\Gamma=0,\quad
\delta_{\text{C}}\Delta=\alpha\Gamma,\quad
\delta_{\text{C}}v^\phi=0,\quad
\delta_{\text{C}}\gamma=0,
\end{equation}
which is the vanishing-$k$ limit of \eqref{locLor-genf}.

Finally, using the modified Lie bracket \eqref{modlie}, the set of generators \eqref{kiladsgencar} acquires the structure of an algebra with composition rules fitting the vanishing-$k$ limit of 
\eqref{AdS-algebra-f}, \eqref{AdS-algebra-Y}, \eqref{AdS-algebra-S}, \eqref{AdS-algebra-Z}:
\begin{eqnarray}
f_3&=&S_1 f_2-S_2f_1+Z_1Y_2-Z_2Y_1+\theta^\ast\left(Y_1f_2-Y_2f_1\right)+\delta_{\xi_1}f_2-\delta_{\xi_2}f_1,
\label{flat-algebra-f}
\\
Y_3&=&S_1 Y_2-S_2Y_1+\theta\left(f_1Y_2-f_2Y_1\right)+\delta_{\xi_1}Y_2-\delta_{\xi_2}Y_1,
\label{flat-algebra-Y}
\\
S_3&=&f_1\upsilon\left(S_2\right)-f_2 \upsilon\left(S_1\right)+Y_1 \ast\upsilon\left(S_2\right)-Y_2 \ast\upsilon\left(S_1\right)+\delta_{\xi_1}S_2-\delta_{\xi_2}S_1,
\label{flat-algebra-S}
\\
Z_3&=&f_1 \upsilon\left(Z_2\right)-f_2 \upsilon\left(Z_1\right)+Y_1 \ast\upsilon\left(Z_2\right)-Y_2 \ast\upsilon\left(Z_1\right)+\delta_{\xi_1}Z_2-\delta_{\xi_2}Z_1.
\label{flat-algebra-Z}
\end{eqnarray}
Observe the different $Y$ composition \eqref{flat-algebra-Y}, compared  to its anti-de Sitter counterpart  \eqref{AdS-algebra-Y}. It reflects the known differences among residual symmetries in flat and anti-de Sitter spacetimes, as e.g. the BMS substituting the Witt algebra. The above result generalizes and unifies previous discussions on this matter \cite{CCMPS,Grumiller:2017sjh, Barnich:2009se}. 

With the achievements collected in the current chapter, we are equipped for comparing the AdS or flat fluid/gravity Eddington--Finkelstein gauge with Bondi or Fefferman--Graham gauges. 


\section{The Bondi gauge and its hydrodynamic frame}\label{BG}

\subsection{Gauge fixing, solution space and residual diffeomorphisms}

In Bondi gauge, the metric takes the form \cite{Bondi1962, Sachs1962-2,Barnich:2010eb}
\begin{equation}
 \label{genbon} 
\text{d}s^2= \text{e}^{2\beta} \frac{V}{r}\text{d}u^2-2 \text{e}^{2\beta}\text{d}u\text{d}r+g \left(\text{d}\phi-U\text{d}u\right)^2.
\end{equation}
Four undetermined functions define a priori the three-dimensional metric: $\beta(u, \phi,r)$, $V(u, \phi,r)$, $g(u, \phi,r)$ and $U(u, \phi,r)$. The gauge-fixing conditions are indeed
\begin{equation}
G_{rr} = 0, \quad G_{r \phi} = 0,
\label{Bondi gauge fixing}
\end{equation} 
plus the determinant condition
\begin{equation}
\partial_r \left(\frac{G_{\phi\phi}}{r^2} \right) =0,
\end{equation}
which leads to $g =  r^2 \text{e}^{2 \varphi}$ with $\varphi$ a function of $(u, \phi)$.

Bondi gauge can be generalized in higher dimensions. In contrast with the fluid/gravity derivative expansion,  it is defined a priori off-shell, and accounts exactly for the local degrees of freedom irrespective of the dynamics. It can be used for finding Einstein or Ricci-flat spacetimes -- or non-vacuum solutions. Generically, Bondi gauge possesses residual diffeomorphisms generated by vectors $\xi(u,\phi,r)= \xi^r \partial_r+\xi^u\partial_u+\xi^\phi \partial_\phi$,  and obtained by requiring (see e.g. \cite{Barnich:2010eb})
\begin{equation}
\mathscr{L}_\xi G_{rr} = 0, \quad \mathscr{L}_\xi G_{r \phi} = 0, \quad \partial_r\left(G^{\phi\phi} \mathscr{L}_\xi G_{\phi \phi} \right)=0 .
\end{equation} 
Solving the latter, we find
\begin{eqnarray}
\xi^u &=& \xi^u_{(0)}, \label{resB1}\\
\xi^\phi &=& \xi^\phi_{(0)}- \text{e}^{-2 \varphi}\partial_\phi \xi^u  \int_r^{+\infty} \frac{\text{d} r'}{{r'}^2}  \text{e}^{2\beta} ,\label{resB2}\\
\xi^r &=& r \left[\omega + U \partial_\phi  \xi^u- \partial_\phi \xi^\phi- \xi^\phi \partial_\phi \varphi -  \xi^u \partial_u \varphi \right],
\label{resB3}
\end{eqnarray} 
where $ \xi^u_{(0)}$, $ \xi^\phi_{(0)}$ and $\omega$ are arbitrary functions of $ (u, \phi)$. It turns out that $G^{\phi\phi} \mathscr{L}_\xi G_{\phi \phi}=2\omega$.

In order to compare \eqref{genbon} with the fluid/gravity expressions \eqref{papaefgenrescrec3d} or \eqref{papaefresricfgen}, we must solve the $rr$, $r\phi$ and $ru$ components of Einstein's equations  with and without cosmological constant. 
Since $G_{rr}=0$ irrespective of $\Lambda=-k^2$, the radial equation is  $R_{rr}=0$. This is solved with  
$\beta = \beta_0 (u, \phi)$.
Similarly, the $r\phi$ equation is $R_{r\phi}=0$ and leads to  
\begin{equation}
\label{U}
U= U_0 + \frac{2}{r}   \text{e}^{2\left(\beta_0-\varphi\right)} \partial_\phi \beta_0  - \frac{N}{r^2}  \text{e}^{2\left(\beta_0-\varphi\right)}
\end{equation} 
with $U_0(u, \phi)$ and $N(u, \phi)$ two arbitrary functions. Finally, the $ru$ equation depends explicitly on $\Lambda$, and so does its general solution, which reads:
\begin{equation}
\label{V}
\frac{V}{r} = -r^2k^2 \text{e}^{2 \beta_0} - 2 r \left(\partial_u \varphi + \partial_\phi U_0 + U_0 \partial_\phi \varphi  \right) + M + \frac{4N}{r}  \text{e}^{2\left(\beta_0-\varphi\right)} \partial_\phi \beta_0 - \frac{N^2}{r^2} \text{e}^{2\left(\beta_0-\varphi\right)} 
\end{equation}
with $M (u, \phi)$ yet another function. In summary, the metric is parameterized by five arbitrary functions of $(u,\phi)$: $\varphi$, $U_0$, $\beta_0$, $M$ and $N$. 

With these results, the Bondi metric \eqref{genbon}, now partly on-shell, fits exactly the fluid/gravity expressions \eqref{papaefgenrescrec3d} or \eqref{papaefresricfgen}, provided the boundary-fluid velocity field is chosen appropriately, namely with $u_\phi$ or $\mu_\phi$ frozen to zero. This defines a specific hydrodynamic frame, which will be referred to as  \emph{Bondi frame}.
\begin{itemize}
\item For relativistic fluids,  $u_\phi=0$ amounts to setting $\Delta=0$ in Eqs. \eqref{genf} and \eqref{genv}. 
Thus, the Bondi-frame velocity forms are
\begin{equation}
\label{bformsads}
\mathbf{u} = -k^2 \frac{\text{d}u}{\gamma}, \quad \ast \mathbf{u} = k\Gamma \left(\text{d}\phi-v^\phi\text{d}u\right).
\end{equation}
These expressions bear three arbitrary boundary functions, i.e. one less than for generic velocity fields. With those, the boundary metric \eqref{genmetr} reads: 
\begin{equation}
\text{d}s^2
=-k^2 \frac{\text{d}u^2}{\gamma^2}
+\Gamma^2
\left(\text{d}\phi-v^\phi\text{d}u\right)^2.
\label{ds2bon}
\end{equation}

\item For Carrollian fluids, the equivalent objects are again expressed in terms of three arbitrary functions, $\Gamma(u,\phi)$, $v^\phi(u,\phi)$ and $\gamma(u,\phi)$: 
\begin{equation}
\label{bformsflat}
\pmb{\mu} = - \frac{\text{d}u}{\gamma}, \quad\ast \pmb{\mu} = \Gamma \left(\text{d}\phi-v^\phi\text{d}u\right).
\end{equation}

The Carrollian degenerate metric \eqref{genmetrc} is now
 \begin{equation}
\text{d}\ell^2
=\Gamma^2
\left(\text{d}\phi-v^\phi\text{d}u\right)^2.
\label{carmetbon}
\end{equation}
\end{itemize}

The fluid/gravity geometric and kinematic data $\Gamma$, $v^\phi$ and $\gamma$ are related to the partly on-shell Bondi functions $\varphi $, $U_0$ and $\beta_0$ as follows:\footnote{Using the expressions for the expansions displayed in footnotes \ref{relexp} and \ref{carexp}, we obtain $
\frac{\Theta^\ast}{k}=\theta^\ast=-\frac{1}{\Gamma}\partial_\phi\ln\gamma=2\text{e}^{-\varphi}\partial_\phi \beta_0
$
and
$
\Theta=\theta=\gamma\left(\partial_u\ln \Gamma
+v^\phi \partial_\phi\ln \Gamma +\partial_\phi v^\phi 
\right)=\text{e}^{-2\beta_0}\left(
\partial_u\varphi
+U_0 \partial_\phi\varphi +\partial_\phi U_0 
\right)
$.
} 
\begin{equation}
\label{dico}
\Gamma
=\text{e}^{\varphi},\quad
v^\phi=U_0,\quad
\gamma=
\text{e}^{-2\beta_0}.
\end{equation}
Finally, the fluid energy and heat densities read:
 \begin{equation}
\label{BDvarep}
8\pi G \varepsilon =\text{e}^{-2\beta_0}M+4\text{e}^{-2\varphi}\left(\partial_\phi \beta_0\right)^2
\end{equation}
and 
\begin{equation}
\label{BDchi}
4\pi G \chi =-k\text{e}^{-\varphi}N \quad \text{or}\quad 4\pi G \zeta=-\text{e}^{-\varphi}N
\end{equation}
for the relativistic or Carrollian situation, respectively. These exhibit the expected relationship between the Bondi mass and the fluid energy density, as well as the remarkable identification of the angular-momentum aspect with the fluid heat density. 

The bulk metrics at hand solve all Einstein's equations, provided the energy and heat densities obey the hydrodynamic equations \eqref{T-cons-el-mag-nc-force} or \eqref{T-cons-el-mag-nc-car}, respectively for the relativistic and Carrollian case. Indeed, the remaining  $uu$, $u\phi$ and $\phi\phi$ equations are the $\mathbf{u}\mathbf{u}$, $\ast\mathbf{u}\ast\mathbf{u}$ and $\ast\mathbf{u}\mathbf{u}$ equations in the anti-de Sitter triad, or the $\pmb{\mu}\pmb{\mu}$, $\ast\pmb{\mu}\ast\pmb{\mu}$ and $\ast\pmb{\mu}\pmb{\mu}$ in the flat counterpart. The hydrodynamic equations can be expressed explicitly in terms of the functions $\varphi $, $U_0$, $\beta_0$, $M$ and $N$, with non-vanishing or vanishing $k$. These are displayed in \eqref{T-cons-el-mag-nc-force-B-en} and \eqref{T-cons-el-mag-nc-force-B-mom}.

In Sec. \ref{genbulk}, we presented a detailed analysis of the residual diffeomorphisms. Irrespective of $\Lambda$, these were generated by four arbitrary functions $f$, $Y$, $S$ and $Z$. For these diffeomorphisms to respect the Bondi hydrodynamic frame, we must require that $\delta_\xi u_\phi=0$ or $\delta_\xi \mu_\phi=0$. Using Eqs. \eqref{deltaudexpl} or \eqref{deltaudexpllim} and imposing $u_\phi=0$ or $\mu_\phi=0$ we find:
\begin{equation}
\label{ZfY}
\psi^\ast=0 \Leftrightarrow Z =\frac{\Theta^\ast}{k} f  -\frac{\ast\text{u}( f )}{k}\quad \text{or}\quad Z =\theta^\ast f  -\ast\upsilon(f),
\end{equation}
in anti de Sitter or Minkowski respectively. 
Using \eqref{dico}, this is recast as
\begin{equation}
\label{ZfY2}
Z =-\frac{1}{\gamma\Gamma}\partial_\phi (\gamma f)
=- \text{e}^{-\varphi}\left(\partial_\phi f-2f\partial_\phi\beta_0\right)
\end{equation}
and the corresponding $\xi$ (Eq. \eqref{kiladsgen} or \eqref{kiladsgencar}) fits the  Bondi residual diffeomorphism generator found in 
\eqref{resB1}, \eqref{resB2}, \eqref{resB3} with
\begin{equation}
\label{Bdiff1}
 \xi^u_{(0)}=\text{e}^{-2\beta_0}f,\quad  \xi^\phi_{(0)}=  \text{e}^{-\varphi}Y+\text{e}^{-2\beta_0}f U_0,
\end{equation}
and
\begin{equation}
\label{Bdiff2}
\omega= S+\Theta f+\frac{\ast\text{u}(Y)}{k}, \quad \text{or}\quad \omega= S+\theta f+\ast\upsilon(Y),
\end{equation}
fitting exactly the $\omega$ introduced in Eqs. \eqref{deltuastud} as an auxiliary function (together with $\omega^\ast$, $\psi$ and $\psi^\ast$).

The effect of these diffeomorphisms on the data defining the solutions ($\Gamma$, $v^\phi$, $\gamma$, $\varepsilon$, and $\chi$ versus $\zeta$,  or $\varphi$, $U_0$, $\beta_0 $, $M$ and $N$) is inferred from the general expressions  \eqref{deltaudexpl}, \eqref{deltastudexpl}, \eqref{deltaeps} and \eqref{deltach}, or  \eqref{deltaudexpllim}, 
\eqref{deltaepscar}  \eqref{deltachcar}, by setting $u_\phi=0$ or $\mu_\phi=0$ and using $Z$ as given in \eqref{ZfY} or \eqref{ZfY2}. We gather these formulas in App. \ref{app1} and carry out the algebra of  Bondi residual diffeomorphisms (depending on three functions, $f$, $Y$ and $S$)  in App.  \ref{app2}.

\subsection{From an arbitrary hydrodynamic frame  to the Bondi frame}

We would like to close this chapter and investigate the class of diffeomorphisms connecting the Bondi frame to arbitrary frames of the fluid/gravity Eddington--Finkelstein gauge. This amounts to restoring (or removing) the sixth function of the solution space ($u_\phi$ or $\mu_\phi$) using some appropriate (and non-unique) combination of the four functions labeling the general residual diffeomorphism. Put differently, this specific combination will parameterize the component $u_\phi$ or $\mu_\phi$.

To this end, we should perform an appropriate residual diffeomorphism, and the most economical  corresponds to a boundary local Lorentz or Carrollian boost. This family has been identified in Sec. \ref{genbulk} as generated by the vectors \eqref{kiladsgenL} or  \eqref{kilmingenC}, labeled by a unique arbitrary function of boundary coordinates, $\psi^\ast$ or $\alpha$. We will provide a unified treatment for anti-de Sitter or Minkowski, using $Z=\nicefrac{\psi^\ast}{k}=\alpha$. 

In the universal parameterization \eqref{genf} or 
\eqref{genfvc}, the form components $\nicefrac{u_\phi}{k}$ or $\mu_\phi$ are identified with $\Delta$, and the effect of the diffeomorphisms under consideration on the function $\Delta$ is precisely a shift (see Eqs. \eqref{locLor-genf} or \eqref{locCar-genfvc}):
\begin{equation}
\label{loc-gen}
\delta\Delta=Z\Gamma.
\end{equation}
Hence, starting from a frame with non-vanishing $\Delta$, i.e. laying outside the Bondi gauge, we can 
reach the latter using a diffeomorphism generated by \eqref{kiladsgenL} or \eqref{kilmingenC} with $\nicefrac{\psi^\ast}{k}=\alpha=Z=\nicefrac{-\Delta}{\Gamma}$, assuming  $\Delta$ small. Keeping the 
lowest $\Delta$-order in the components of the generating vector field, we find the following coordinate transformation:
\begin{eqnarray}
r_{\text{B}}&=&r+\left.\xi^r\right\vert_{\mathcal{O}(\Delta)}+\mathcal{O}\left(\Delta^2\right)=r-\frac{\Delta}{\Gamma^2}\partial_\phi\ln \gamma
+\frac{1}{\Gamma}\partial_\phi\frac{\Delta}{\Gamma}
+\frac{4\pi G \Delta}{r\Gamma}\zeta+\mathcal{O}\left(\Delta^2\right)
,\label{anytoBr}
\\
u_{\text{B}}&=&u+\left.\xi^u\right\vert_{\mathcal{O}(\Delta)}+\mathcal{O}\left(\Delta^2\right)=u+\mathcal{O}\left(\Delta^2\right),\label{anytoBu}
\\
\phi_{\text{B}}&=&\phi+\left.\xi^\phi\right\vert_{\mathcal{O}(\Delta)}+\mathcal{O}\left(\Delta^2\right)=\phi -\frac{\Delta}{r\Gamma^2}+\mathcal{O}\left(\Delta^2\right).
\label{anytoBphi}
\end{eqnarray}
We have dropped the index $\text{L}$ or  $\text{C}$ in the generating vectors  \eqref{kiladsgenL} or \eqref{kilmingenC} because, expressed in terms of $\Delta$ (and $\zeta$ instead of $\chi$), the latter do not depend on $k$. The infinitesimal coordinate transformation  \eqref{anytoBr}, \eqref{anytoBu},   \eqref{anytoBphi} relating the Bondi frame to some more general hydrodynamic frame is the same, irrespective of the bulk being AdS or Minkowski.  However, due to the explicit $k$-dependence of the variations produced by this generator (see \eqref{deltaudexpl}, \eqref{deltastudexpl}, \eqref{deltaeps}, \eqref{deltach}), their integrated, finite transformations will depend on $k$, as we will shortly see. 

In order to treat situations with arbitrary $\Delta$, i.e. initially far from the Bondi gauge, one
should integrate the orbits of the above generating vector fields. This can be performed in a series expansion in powers of $\nicefrac{1}{r}$, and at lowest order the diffeomorphism leads to
\begin{eqnarray}
r_{\text{B}}-r&=&
\frac{\gamma}{\Gamma^2}
\left[
\frac{\Delta}{\gamma}\partial_\phi\ln \frac{\Delta}{\gamma\Gamma}
+\frac{v^\phi}{k^2K+1}\left(\Delta\partial_\phi\Delta-
K \Gamma\partial_\phi
\Gamma
\right)
+\frac{\partial_\phi v^\phi}{k^2K+1}\left(\Delta^2-K\Gamma^2\right)\right.
\nonumber
\\
&&\left.+\frac{1}{k^2K+1}\left(\Delta\partial_u\Delta-
K \Gamma\partial_u
\Gamma
\right)\bigg]+\mathcal{O}(\nicefrac{1}{r})\right]
,
\\
u_{\text{B}}-u&=&-\frac{\gamma K}{r}+\mathcal{O}(\nicefrac{1}{r^2})
\end{eqnarray}
and 
\begin{eqnarray}
\phi_{\text{B}}-\phi&=&-\frac{1}{r}\left(
\frac{\Delta}{\Gamma^2}
+ \gamma v^\phi K\right)
+\mathcal{O}(\nicefrac{1}{r^2}).
\end{eqnarray}
The function $K(u,\phi)$ is defined  as
\begin{equation}
K=\frac{1}{k^2}\sqrt{1+\frac{k^2\Delta^2}{\Gamma^2}}-\frac{1}{k^2},
\end{equation}
and the coordinate transformation at hand is regular in the flat limit thanks to 
\begin{equation}
\lim_{k\to 0}K
=\frac{\Delta^2}{2\Gamma^2}.
\end{equation}
The functions $\varepsilon$ and $\chi$ (or $\zeta$) will appear at higher order in the $\nicefrac{1}{r}$ expansion. 


\section{The Fefferman--Graham gauge for anti de Sitter}\label{FGG}

The mathematical foundations of holography rely on the existence of the Fefferman--Graham expansion for asymptotically anti-de Sitter Einstein spacetimes \cite{PMP-FG1, PMP-FG2}. This expansion is based on the following ansatz, which defines the Fefferman--Graham gauge:
\begin{equation}
\text{d} s^2 = \frac{\text{d} \rho^2}{k^2\rho^2}  + G_{\alpha\beta}(\rho, x) \text{d} x^\alpha \text{d} x^\beta,
\label{gfFG}
\end{equation}
reflecting the gauge-fixing conditions 
\begin{equation}
G_{\rho \rho} = \frac{1}{k^2\rho^2}, \quad G_{\rho \beta} = 0.
\label{gfFG fixing}
\end{equation}
As for the Bondi gauge, there are residual gauge diffeomorphisms, which preserve the gauge-fixing conditions \eqref{gfFG fixing}. Their generators satisfy
\begin{equation}
\mathscr{L}_\xi G_{\rho \rho} = 0, \quad \mathscr{L}_\xi G_{\rho \beta} = 0.
\end{equation}
The explicit solutions to these equations are given by 
\begin{equation}
\xi^\rho =  \rho \sigma, \quad \xi^\alpha = \xi^\alpha_{(0)} - \frac{1}{k^2} \partial_\beta  \sigma \int^\rho_0 \frac{\text{d} \rho'}{\rho'} G^{\alpha\beta} (\rho', x), 
\label{residual gauge diffeomorphisms FG}
\end{equation}
where $\sigma$ and $\xi^\alpha_{(0)}$ are three arbitrary functions of $x^\alpha$.

In the gauge at hand, one can find the general three-dimensional solution to Einstein's equations in the form of a truncated series expansion in powers of $\rho^2$. We impose the preliminary boundary condition $G_{\alpha\beta} = \mathcal{O}(\nicefrac{1}{\rho^2})$ -- the conformal boundary is located at $\rho=0$. Solving the $\rho\rho$ and $\rho \beta$ Einstein's equations we obtain
\begin{equation}
G_{\alpha\beta}(\rho , x)= \rho^{-2} G_{\alpha\beta}^{(0)}(x) +  G_{\alpha\beta}^{(2)}(x)  + \rho^2  G_{\alpha\beta}^{(4)}(x),
\label{part-on-shell}
\end{equation} 
where $G_{\alpha\beta}^{(4)}$ is determined by $G_{\alpha\beta}^{(0)}$ and $G_{\alpha\beta}^{(2)}$ as
\begin{equation}
 G_{\alpha\beta}^{(4)} = \frac{1}{4} G^{(2)}_{\alpha\gamma}G_{(0)}^{\gamma\delta}G^{(2)}_{\delta\beta}.
\label{part-on-shell2}
\end{equation} 
This simple computation illustrates the general 
Fefferman--Graham ambient metric construction \cite{Anderson:2004yi}, according to which 
asymptotically locally anti-de Sitter spacetimes are determined by a set of independent boundary data,
here $G_{\alpha\beta}^{(0)}$ and $G_{\alpha\beta}^{(2)}$. 

The first piece, $G_{\alpha\beta}^{(0)}$, is interpreted as the boundary metric, generically spelled $g_{\alpha\beta}$. It is not subject to any conditions. The second,  $G_{\alpha\beta}^{(2)}$, appears as a boundary tensor, the trace of which, computed with the boundary metric, must be proportional to the Riemann--Christoffel scalar of the latter: $\text{Tr}\left[G^{(2)}\right]=-\frac{R}{2k^2}$. It is related to the holographic energy--momentum tensor  \cite{Balasubramanian:1999re, Skenderis2001} as\footnote{We normalize the energy--momentum tensor with an extra factor $2k$ compared to the quoted literature. This heterodox choice is made to comply with the normalizations used in the fluid/gravity approach of Sec. \ref{FGDE}, based on the standard relativistic-fluid energy--momentum tensor \eqref{Tgen}.} 
\begin{equation}
\tilde T_{\alpha\beta} = \frac{k^2}{4 \pi G} \left(G^{(2)}_{\alpha\beta} - g_{\alpha\beta}\text{Tr} \left[G^{(2)}\right]
\right).
\label{trace condition}
\end{equation}
This tensor is tracefull with $\tilde T^\alpha_{\hphantom{\alpha}\alpha}=\frac{kc}{12\pi}R$, where  $c = \nicefrac{3 }{2k G}$ is the three-dimensional Brown--Henneaux central charge \cite{Brown:1986nw, Henningson1998}. In particular, the remaining $\alpha\beta$ field equations translate into the covariant conservation of $\tilde T_{\alpha\beta}$ with the boundary Levi--Civita connection: $\nabla^\alpha\tilde T_{\alpha\beta}=0$. This is a set of differential equations for the boundary data, encoded in five arbitrary functions.

Owing to the solution \eqref{part-on-shell}, we can reconsider the residual diffeomorphisms of the gauge at hand, Eq. \eqref{residual gauge diffeomorphisms FG}. These are given in a closed form, where the $\rho$ dependence is explicit. Expanding in powers of $\rho$ we find
\begin{equation}
\xi^\rho = \sigma \rho , \quad \xi^\alpha = \xi^\alpha_{(0)} - \frac{k^2\rho^2}{2} g^{\alpha\beta}  \partial_\beta  \sigma + \frac{k^2\rho^4}{4} g^{\alpha \gamma}G_{\gamma \delta}^{(2)}g^{\delta \beta} \partial_\beta  \sigma + \mathcal{O}\left(\rho^6\right) .
\label{FGdiffs}
\end{equation}
Under these residual gauge diffeomorphisms, the unconstrained part of the solution space transforms as 
\begin{equation}
\delta_\xi g_{\alpha\beta}= -\mathscr{L}_{\xi_{(0)}} g_{\alpha\beta}+ 2 \sigma g_{\alpha\beta},
\label{varmetric}
\end{equation} 
while the constrained part transforms as
\begin{equation}
\delta_\xi G_{\alpha\beta}^{(2)} = - \mathscr{L}_{\xi_{(0)}} G_{\alpha\beta}^{(2)} + \frac{1}{2k^2} \mathscr{L}_{\partial \sigma} g_{\alpha\beta}
\label{varem}
\end{equation}
with $\partial \sigma=g^{\alpha\beta}\partial_\beta\sigma \partial_\alpha$. Equation \eqref{varmetric} provides the transformation of the boundary metric, whereas the variation of the energy--momentum tensor \eqref{trace condition} can be extracted from \eqref{varem}.

For the sake of completeness, we would like to mention a further restriction on the boundary, often called in the literature Brown--Henneaux condition\cite{Brown:1986nw}: the boundary metric is frozen to be the flat metric $g_{\mu\nu}=\eta_{\mu\nu}$. When imposing this condition we recover the usual asymptotic symmetry group in AdS$_3$, i.e. the conformal group. Indeed, \eqref{varmetric} becomes
\begin{equation}
\delta_\xi \eta_{\mu\nu} = -\mathscr{L}_{\xi_{(0)}} \eta_{\mu\nu} + 2 \sigma \eta_{\mu\nu}=0.
\end{equation}  
Therefore the symmetry algebra is uniquely specified by a boundary vector $\xi_{(0)}$ that belongs to the boundary conformal algebra. Using this phase space, one can compute the associated surface charges and their algebra, and deduce the Brown--Henneaux central charge.

We would like to end this overview on the Fefferman--Graham approach, and relate it to the Bondi gauge discussed in Sec. \ref{BG}. In both gauges, the solution space is spanned by five functions, obeying a system of  two conservation equations. Suppose we start with a locally AdS spacetime in Bondi gauge, coordinatized with $(r, u, \phi)$ and labelled by $\varphi(u,\phi)$, $U_0(u,\phi)$ and $\beta_0(u,\phi)$, as well as $M(u,\phi)$ and $N(u,\phi)$. What is the explicit diffeomorphism that achieves the mapping of the spacetime under consideration onto the Fefferman--Graham gauge, and  what are the fundamental data labelling the solution in this gauge? 

In order to answer these questions, we follow \cite{Poole:2018koa, Compere:2019bua} and proceed in two steps. We firstly move from Bondi to tortoise coordinates $(r,u,\phi) \to (r^*,t^*,\phi^*)$:
\begin{equation}
kr = -  \cot \left(kr^*\right), \quad u = t^* - r^*, \quad \phi^* = \phi. 
\label{BT}
\end{equation}
Fefferman--Graham coordinates $(\rho,t,\vartheta)$ are reached in the second step as series expansions:
\begin{eqnarray}
r^* &=& R_1(t,\vartheta) \rho + R_2(t,\vartheta)\rho^2 + R_3(t, \vartheta) \rho^3 + \mathcal{O}\left(\rho^4\right),
\label{diffeo-Bondi-FG2}
\\
t^* &=& t + T_1(t,\vartheta) \rho +  T_2(t,\vartheta) \rho^2 + T_3 (t, \vartheta) \rho^3 + \mathcal{O}\left(\rho^4\right), 
\label{diffeo-Bondi-FG1}
 \\
\vartheta^* &=& \vartheta + \Theta_1(t,\vartheta) \rho + \Theta_2 (t,\vartheta) \rho^2 + \Theta_3 (t, \vartheta) \rho^3 + \mathcal{O}\left(\rho^4\right).
\label{diffeo-Bondi-FG3}
\end{eqnarray} 
The functions $T_i (t,\vartheta)$, $R_i(t,\vartheta)$ and $\Theta_i (t,\vartheta)$ can be worked out explicitly. For the sake of brevity, we report here only the first orders:
\begin{eqnarray}
&&\begin{cases}
R_1(t, \vartheta) = \frac{1}{k^2},\\
R_2 (t, \vartheta)= -  \frac{1}{k^4}\text{e}^{-2\beta_0}\left(\partial_\phi U_0 + U_0 \partial_\phi \varphi + \partial_t \varphi \right), 
\end{cases}
\\
&&\begin{cases}
T_1(t, \vartheta) =\frac{1}{k^2} \left(1 - \text{e}^{-2\beta_0}\right),\\
T_2 (t, \vartheta)= -  \frac{1}{k^4}\text{e}^{-4\beta_0} \left[\text{e}^{2\beta_0}\partial_\phi U_0 + U_0 \left(\partial_\phi\beta_0 + \text{e}^{2\beta_0} \partial_\phi \varphi\right) + \partial_t \beta_0 + \text{e}^{2\beta_0} \partial_t \varphi \right]
\end{cases}
\end{eqnarray}
and
\begin{eqnarray}
&&\begin{cases}
\Theta_1(t, \vartheta) = -\frac{1}{k^2} \text{e}^{-2\beta_0} U_0,\\
\Theta_2(t, \vartheta)=  \frac{1}{2k^2} \left[2 \text{e}^{-2\varphi} \partial_\phi \beta_0  + \frac{1}{k^2}\text{e}^{-4\beta_0}\left( \partial_t U_0+ U_0 \left(\partial_\phi U_0 - 2 \partial_t \beta_0\right)- 2  U_0^2 \partial_\phi \beta_0\right)\right]. 
\end{cases}
\end{eqnarray}
In all these expressions, the functions  $\varphi$, $U_0$ and $\beta_0$ have arguments $(t,\vartheta)$. 

The transformations \eqref{BT}, \eqref{diffeo-Bondi-FG2}, \eqref{diffeo-Bondi-FG1}, \eqref{diffeo-Bondi-FG3}, map the metric \eqref{genbon} with \eqref{U} and \eqref{V} onto \eqref{gfFG}. In the latter Fefferman--Graham form, we read off the boundary data $G_{\alpha\beta}^{(0)}$ and $G_{\alpha\beta}^{(2)}$, equivalently cast as boundary metric $g_{\alpha\beta}$ and boundary energy--momentum tensor  $\tilde T_{\alpha\beta}$. We find  
\begin{equation}
g_{\alpha \beta}\text{d}x^\alpha\text{d}x^\beta  
=-k^2\text{e}^{4\beta_0}\text{d}t^2
+\text{e}^{2\varphi}
\left(\text{d}\vartheta-U_0 \text{d}t\right)^2,
\label{ds2bonFG}
\end{equation}
which is precisely \eqref{ds2bonexpl} with $(u,\phi)$ replaced by $(t,\vartheta)$,
whereas $\tilde T_{tt}$,  $\tilde T_{t\vartheta}$ and  $\tilde T_{\vartheta\vartheta}$ coincide with 
$\tilde T_{uu}$,  $\tilde T_{u\phi}$ and  $\tilde T_{\phi\phi}$ displayed in \eqref{Tuu}, \eqref{Tuphi}, \eqref{Tphiphi} with all arguments $(u,\phi)$ traded for $(t,\vartheta)$.  In the same  vein, the transformations \eqref{BT}, \eqref{diffeo-Bondi-FG2}, \eqref{diffeo-Bondi-FG1}, \eqref{diffeo-Bondi-FG3}, map the Bondi residual diffeomorphisms \eqref{Bdiff1}, \eqref{Bdiff2} onto 
the residual gauge diffeomorphisms found in \eqref{FGdiffs} for the Fefferman--Graham gauge with
\begin{equation}\label{map}
\begin{split}
&\xi_{(0)}^t = \text{e}^{-2\beta_0}f, \quad
\xi_{(0)}^\vartheta =  \text{e}^{-\varphi}Y+\text{e}^{-2\beta_0}f U_0, \\
&\sigma =- \omega  + \text{e}^{-\varphi}\partial_\vartheta Y +\text{e}^{-2\beta_0}f
\left(\partial_t \varphi+U_0 \partial_\vartheta \varphi+\partial_\vartheta  U_0
\right)=-S.
\end{split}
\end{equation}
The latter result is valid at leading order only, because the residual-diffeomorphism generators are field-dependent. 
In order to achieve the exact mapping, one has to take this feature into account \cite{Compere:2016hzt}.

These results do not come as a surprise. The conformal boundary is located at $\rho=0$ in the Fefferman--Graham gauge, which corresponds to infinite  $r$ according to the above diffeomorphism. This is in agreement with the analysis performed in the fluid/gravity side (Sec.~\ref{FGDE}), which includes the Bondi gauge (Sec. \ref{BG}). On the conformal boundary, the Fefferman--Graham/Bondi coordinate transformation trivializes (see \eqref{BT}, \eqref{diffeo-Bondi-FG2}, \eqref{diffeo-Bondi-FG1}, \eqref{diffeo-Bondi-FG3}): $u=t$ and $\phi=\vartheta$. Hence, the boundary metric and the holographic energy--momentum tensor remain unaltered. This observation does not downplay the present analysis. As already stressed in Sec. \eqref{genbulk}, although the holographic reconstruction of the bulk from the boundary is successfully conducted in the Eddington--Finkelstein gauge of fluid/gravity correspondence, the holographic energy--momentum tensor cannot be singled out in the derivative expansion -- in dimension higher than three it is actually scattered in many orders in the $\nicefrac{1}{r}$ series. The information is present though, and our observation here is that it can be recomposed by an appropriate diffeomorphism, which does not alter the boundary. The radial coordinate is space-like for Fefferman--Graham and light-like in Bondi gauge (and in more general hydrodynamic frames of the fluid/gravity correspondence), but their complicated relationship simplifies on the boundary.

\section{Conclusion}

The salient features of the present work come as our response to two basic questions. What is the solution space of the fluid/gravity derivative expansion viewed as a gauge, and what are its residual diffeomorphisms? Where do the conventional Bondi and Fefferman--Graham gauges stand regarding  fluid/gravity, and how is this web woven?  

The fluid/gravity derivative expansion is a sort of Eddington--Finkelstein gauge, where the anti-de Sitter or flat bulk spacetimes are described in terms of six functions. These account for all boundary data: geometry, fluid kinematics and fluid dynamics -- relativistic or Carrollian, i.e. for AdS or Minkowski, respectively. These data obey two evolution equations, the energy and momentum conservation laws, which feature the two transverse Einstein's bulk equations. 

The set of residual diffeomorphisms  is generated by bulk vectors depending on four arbitrary functions. These parameterize the variations produced in the bulk, reflecting faithfully the variations of the boundary data, and mapping on-shell fluid configurations onto each other. Of these transformations, one has a special status. It generates boundary local  boosts (Lorentz or Carrollian), which leave the geometry as well as the fluid dynamics invariant, but modify its kinematics. This is the well-known hydrodynamic-frame freedom, generically valid for relativistic fluids, which occasionally survives in specific Carrollian or even Galilean instances. 

One of the six defining functions can be used to lock the boundary-fluid velocity. Equivalently one of the four residual diffeomorphisms provides the designated tool for changing hydrodynamic frame. Remarkably, the Bondi gauge is part of the fluid/gravity landscape with the mass identified to the fluid energy density, and the angular momentum  to the heat density. It 
corresponds to a specific hydrodynamic frame we have named Bondi frame. This explains why, from the fluid side, the Bondi solution space is supported by five arbitrary functions (obeying two equations), while it is form-invariant under a three-parameter family of
diffeomorphisms.  It also makes the relationship of the Bondi to the general fluid/gravity gauge straightforward: every general fluid/gravity configuration is amenable to Bondi frame using a diffeomorphism associated with a boundary local boost. Integrating the generating vector into a finite transformation is challenging, and should be considered as a noticeable achievement of the present work. 

The Fefferman--Graham gauge stands somehow aside. It is valid exclusively in anti de Sitter and its holographic coordinate is space-like. The holographic data, metric and energy--momentum tensor, appear at precise leading and subleading orders. This is a key feature of this gauge, to be opposed to what happens in Bondi or more generally in the fluid/gravity derivative expansion, where the information stored in the energy--momentum tensor is spread in various orders of the radial expansion. In the Fefferman--Graham gauge the fluid is immaterial, no quantity
such as a fluid velocity exists, and the solution space is described by five arbitrary functions satisfying two equations.  It has residual diffeomorphisms generated by three functions, as in Bondi gauge. Our original contribution was to uncover explicitly the diffeomorphism relating them, mapping the fluid boundary data in the Bondi frame onto the Fefferman--Graham holographic metric and energy--momentum tensor. 

This work sets the stage for further investigation in several directions. The most straightforward and appealing part of this program is definitely the determination of the phase space and asymptotic charges. We unravelled here the algebra obeyed by the residual diffeomorphisms of the fluid/gravity solution space. This is the first step towards a complete understanding of asymptotic symmetries,\footnote{Although performed in a unified fashion, the flat and anti-de Sitter cases are different with this respect. Comparing \eqref{AdS-algebra-Y} and \eqref{flat-algebra-Y} reveals this sharp distinction.} which are expected to be richer than the double copy of Virasoro or $\mathfrak{bms}_3$ appearing in Feffermann--Graham or Bondi gauge. This fact has  been demonstrated in \cite{CCMPS} with a circumscribed determination of charges, sufficient though to unveil the role of the hydrodynamic-frame change, pointing towards an improper gauge transformation.\footnote{From the perspective of relativistic fluid dynamics, this suggests that the hydrodynamic-frame invariance, as discussed in textbooks, is at best a local property.} It echoes some previous works stipulating that a complete gauge fixing should definitely eliminate this sort of transformations \cite{Grumiller:2016pqb , Grumiller:2017sjh}. We expect more general algebras to be concretely realized in the complete three-dimensional flat and anti-de Sitter spacetimes discussed here within the fluid/gravity gauge, as those anticipated in \cite{Troessaert:2013fma,Perez:2016vqo,Grumiller:2016pqb , Grumiller:2017sjh,oblak}.

Extending the present approach to higher dimensions is important and challenging. At the first place, the fluid/gravity bulk reconstruction is based on a genuine expansion, except when special assumptions are made, which guarantee its resummation \cite{CMPPS2}. Secondly, contrary to what happens in three dimensions, the Bondi gauge is intersecting with (rather than embedded in) the fluid/gravity Eddington--Finkelstein framing. Thirdly, the group of Lorentz boosts has more than one parameter. All this disports a severe accretion of difficulty, which should not be discouraging since translating Bondi into fluid data, including Ricci-flat and anti-de Sitter backgrounds, is expected to be physically rewarding. 

\section*{Acknowledgements}

We would like to thank F. Alessio, G. Barnich, A. Campoleoni, G. Comp\`ere, L. Donnay, A.~Fiorucci, G. Giribet, V. Godet, D. Grumiller, R. Leigh, P. Mao, R. Olea, A. Petkou, K. Siampos and C. Zwikel for insightful discussions.  The work of L. Ciambelli is supported by the ERC Advanced Grant \textsl{High-Spin-Grav}. Romain Ruzziconi is a FRIA Research Fellow of the Fonds de la Recherche Scientifique F.R.S.-FNRS (Belgium). We thank each other institutions for hospitality and financial support. Luca Ciambelli thanks Universidad de Buenos Aires and Universidad Andr\'es Bello, where this work was presented, for the warm hospitality. Charles Marteau would like to thank the Kavli Institute for their hospitality and support while part of this work was completed. This research was supported in part by the National Science Foundation under Grant No. NSF PHY-1748958, by the Heising--Simons Foundation and by the ANR-16-CE31-0004 contract \textsl{Black-dS-String}.

\appendix

\section{Solutions, diffeomorphisms and variations in Bondi gauge}
\label{app1}
We assemble in this appendix some explicit formulas in Bondi gauge. These include the solution space and the variations induced by residual diffeomorphisms. 

The solution space of locally anti-de Sitter or locally Minkowski three-dimensional spacetimes in Bondi gauge \eqref{genbon} is parameterized by five functions $\Gamma$, $v^\phi$, $\gamma$, $\varepsilon$, and $\chi$ or $\zeta$, in the boundary-fluid language, or equivalently by $\varphi$, $U_0$, $\beta_0 $, $M$ and $N$. The map between these two analogue sets is given in Eqs. \eqref{dico}, \eqref{BDvarep} and \eqref{BDchi}. The functions  $\varepsilon$ and $\chi$ obey the hydrodynamic equations  \eqref{T-cons-el-mag-nc-force} or \eqref{T-cons-el-mag-nc-car}, which can be recast for $M(u,\phi)$ and $N(u,\phi)$ as
\beqn
\partial_u M &=&- 2 M \partial_u \varphi + 2 M \partial_u \beta_0-2 M\pa_\phi U_0+2 MU_0 \partial_\phi \beta_0- 2 MU_0 \partial_\phi \varphi-U_0\partial_\phi M \nonumber \\
&&+ 2 k^2 \text{e}^{4 \beta_0-2 \varphi} \left[\partial_\phi N + N \left(4 \partial_\phi \beta_0 - \partial_\phi \varphi \right) \right]-2 \text{e}^{2 \beta_0-2 \varphi}\times \nonumber\\
&&\times \left\{\partial_\phi U_0  \left[8  \left(\partial_\phi \beta_0\right)^2 - 4 \partial_\phi \varphi \partial_\phi \beta_0 +  \left(\partial_\phi \varphi\right)^2 + 4  \partial_\phi^2  \beta_0 - 2 \partial_\phi^2 \varphi \right] -\partial_\phi^3 U_0\right. \nonumber\\
&&+ U_0 \left[ \partial_\phi \beta_0  \left(8 \partial_\phi^2 \beta_0 - 2 \partial_\phi^2 \varphi \right) + \partial_\phi \varphi  \left( \partial_\phi^2 \varphi- 2 \partial_\phi^2 \beta_0  \right) + 2 \partial_\phi^3 \beta_0 - \partial_\phi^3 \varphi \right] \nonumber\\
&& \left.+ 2 \partial_u \partial_\phi \beta_0  \left(4 \partial_\phi \beta_0 - \partial_\phi \varphi \right) + \partial_u \partial_\phi \varphi  \left( \partial_\phi \varphi -2 \partial_\phi \beta_0 \right) + 2 \partial_u \partial_\phi^2 \beta_0 - \partial_u \partial_\phi^2 \varphi \right\}
\label{T-cons-el-mag-nc-force-B-en}
\eeqn
and
\beqn
\partial_uN +  N \partial_u \varphi &=& \frac{1}{2} \partial_\phi M+ M\partial_\phi \beta_0   -2 N \partial_\phi U_0 - U_0 \left(\partial_\phi N + N \partial_\phi \varphi \right)\nonumber\\
&& + 4 \text{e}^{2 \beta_0-2 \varphi} \left[2 \left(\partial_\phi \beta_0\right)^3 - \partial_\phi \varphi \left(\partial_\phi \beta_0\right)^2 + \partial_\phi \beta_0  \left(\partial_\phi^2 \beta_0\right) \right].
\label{T-cons-el-mag-nc-force-B-mom}
\eeqn
Equations \eqref{T-cons-el-mag-nc-force-B-en} and \eqref{T-cons-el-mag-nc-force-B-mom} coincide with \eqref{T-cons-el-mag-nc-force} for generic $k$, corresponding to bulk anti-de Sitter spacetimes i.e.  relativistic holographic fluids. These are defined on the conformal boundary equipped with metric \eqref{ds2bon}
\begin{equation}
\text{d}s^2=-k^2\text{e}^{4\beta_0}\text{d}u^2
+\text{e}^{2\varphi}
\left(\text{d}\phi-U_0 \text{d}u\right)^2,
\label{ds2bonexpl}
\end{equation}
and their dynamics \eqref{T-cons-el-mag-nc-force-B-en} and \eqref{T-cons-el-mag-nc-force-B-mom} 
qualifies the conservation of $\tilde{\text{T}}$ defined in \eqref{T-tilde-em}. We expicitely display the latter, putting together all available information (\eqref{Tgen} and \eqref{D}, \eqref{bformsads}, \eqref{dico}, \eqref{BDvarep} and \eqref{BDchi}):
\beqn
\tilde T_{uu} &=& \frac{1}{8\pi G} \text{e}^{-4\beta_0 - 2 \varphi} 
\left\{
4k^2\text{e}^{8 \beta_0} \left[2 \left(\partial_\phi \beta_0\right)^2 - \partial_\phi \beta_0 \partial_\phi \varphi + \partial_\phi^2 \beta_0 \right] 
\right.
\nonumber\\
&&+ \text{e}^{4 \beta_0 + 2 \varphi}\left[k^2 \text{e}^{2 \beta_0} \left(M-4 N U_0 \right)- 
\left( \left(\partial_\phi U_0\right)^2 + U_0^2 \left(4 \partial_\phi^2 \varphi-8 \partial_\phi \beta_0 \partial_\phi \varphi + \left(\partial_\phi \varphi\right)^2 \right) \right.\right.
\nonumber\\
&&
+ \left(\partial_t \varphi\right)^2+ 2 \partial_\phi U_0 \left(U_0 \left(3 \partial_\phi \varphi -4 \partial_\phi \beta_0 \right)+ \partial_t \varphi\right)
\nonumber\\
&&\left.\left.+ 2 U_0 \left(2 \partial_\phi^2 U_0 + \left( \partial_\phi \varphi-4 \partial_\phi\beta_0 \right) \partial_t \varphi + 2 \partial_t \partial_\phi \varphi \right)\right)\right]
\nonumber\\
&&
+  \text{e}^{4 \varphi} U_0^2 \left[ \text{e}^{2\beta_0} M + k^{-2} \left( \left(\partial_\phi U_0\right)^2 +U_0^2 \left( 2 \partial_\phi^2 \varphi- 4 \partial_\phi \beta_0 \partial_\phi \varphi + \left(\partial_\phi \varphi\right)^2  \right)\right.\right.  \nonumber\\
&&  +2 \partial_\phi \varphi \partial_t U_0 + \partial_t \varphi ( \partial_t \varphi-4 \partial_t \beta_0  ) + 2 \partial_\phi U_0 \left(2 U_0 \left( \partial_\phi \varphi - \partial_\phi \beta_0 \right)- 2 \partial_t \beta_0+ \partial_t \varphi \right)  \nonumber \\
&&+ 2 U_0\left(\partial_\phi^2 U_0 - 2 \partial_\phi \beta_0 \partial_t \varphi + \partial_\phi \varphi \left( \partial_t \varphi  -2 \partial_t \beta_0 \right) + 2 \partial_t \partial_\phi \varphi\right) 
\nonumber \\
&&\left. \left. 
+ 2 \left(\partial_t \partial_\phi U_0 + \partial_t^2 \varphi \right)  \Big) \right] \right\},
\label{Tuu}
\\
\tilde T_{u \phi} &=& \frac{1}{8\pi G}\text{e}^{-4\beta_0}\left\{ 2k^2\text{e}^{6 \beta_0} N - 2  \text{e}^{4\beta_0} \Big[\partial_\phi U_0 \left(2 \partial_\phi \beta_0 - \partial_\phi \varphi\right) - \partial_\phi^2 U_0  \right.\nonumber \\
&&\left.+ U_0 \left(2 \partial_\phi \beta_0 \partial_\phi \varphi - \partial_\phi^2 \varphi \right)+ 2 \partial_\phi \beta_0  \partial_t \varphi- \partial_t \partial_\phi \varphi \right] 
\nonumber \\
&&
- \text{e}^{2 \varphi} U_0 \left[  \text{e}^{2 \beta_0} M + k^{-2} \Big( \left(\partial_\phi U_0\right)^2 + U_0^2 \left(2 \partial^2_\phi \varphi-4\partial_\phi \beta_0 \partial_\phi \varphi + \left(\partial_\phi \varphi \right)^2 \right) \right.\nonumber\\
&&+ 2 \partial_\phi \varphi \partial_t U_0 + \partial_t \varphi \left( \partial_t \varphi -4 \partial_t \beta_0 \right) + 2 \partial_\phi U_0 \left( 2 U_0 \left( \partial_\phi \varphi -\partial_\phi \beta_0 \right) -2 \partial_t \beta_0 + \partial_t \varphi\right) \nonumber\\
&&+ 2 U_0 \left(\partial_\phi^2 U_0 - 2 \partial_\phi \beta_0 \partial_t \varphi + \partial_\phi \varphi \left(\partial_t \varphi- 2 \partial_t \beta_0 \right) +2 \partial_t \partial_\phi \varphi\right) 
 \nonumber\\
&&\left.\left.
+ 2 \left(\partial_t \partial_\phi U_0 + \partial_t^2 \varphi \right)\Big)\right] \right\} ,
\label{Tuphi}
\eeqn
and
\beqn
\tilde T_{\phi \phi} &=& \frac{1}{8\pi G} \text{e}^{-4 \beta_0+ 2 \varphi} \left\{ \text{e}^{2 \beta_0}M + k^{-2} \left[ \left(\partial_\phi U_0\right)^2 + U_0^2 \left(2 \partial_\phi^2 \varphi -4 \partial_\phi \beta_0 \partial_\phi \varphi + \left(\partial_\phi \varphi\right)^2 \right)\right.\right.\nonumber \\
&&+ 2 \partial_\phi \varphi \partial_t U_0+ \partial_t \varphi \left(\partial_t \varphi -4 \partial_t \beta_0\right) 
+ 2 \partial_\phi U_0 \left(2 U_0 \left( \partial_\phi \varphi-\partial_\phi \beta_0 \right) - 2 \partial_t \beta_0 + \partial_t \varphi \right) 
\nonumber \\
&&
+ 2 U_0 \left(\partial_\phi^2 U_0 - 2 \partial_\phi \beta_0 \partial_t \varphi +\partial_\phi \varphi \left( \partial_t \varphi-2 \partial_t \beta_0\right) + 2 \partial_t \partial_\phi \varphi \right) \nonumber \\
&&\left.+ 2 \left(\partial_t \partial_\phi U_0 + \partial_t^2 \varphi \right)\Big] \right\}.
\label{Tphiphi}
\eeqn

In the bulk flat limit, reached at boundary velocity of light $k=0$, Eqs. \eqref{T-cons-el-mag-nc-force-B-en} and \eqref{T-cons-el-mag-nc-force-B-mom} fit \eqref{T-cons-el-mag-nc-car}. They describe the hydrodynamics of Carrollian fluids defined on a boundary equipped with the degenerate metric 
\eqref{carmetbon}:
\begin{equation}
\text{d}\ell^2=\text{e}^{2\varphi}
\left(\text{d}\phi-U_0 \text{d}u\right)^2.
\label{carmetbonexpl}
\end{equation}

In Bondi gauge, the residual diffeomorphisms are generated by a vector $\xi$ with components displayed in \eqref{resB3}, 
$U$ given in \eqref{U} and \eqref{Bdiff1} in terms of three arbitrary functions $f$, $Y$ and $\omega$. The effect of these diffeomorphisms on the above data ($\varphi$, $U_0$, $\beta_0 $, $M$ and $N$) defining the solutions is inferred from the general expressions  \eqref{deltaudexpl}, \eqref{deltastudexpl}, \eqref{deltaeps} and \eqref{deltach}, or  \eqref{deltaudexpllim}, 
\eqref{deltaepscar}  \eqref{deltachcar}, by setting $u_\phi=0$ or $\mu_\phi=0$ and using $Z$ as given in \eqref{ZfY} or \eqref{ZfY2}. We find:
\begin{eqnarray}
\delta_\xi\varphi&=& -\omega, 
\\
\delta_\xi U_0&=& k^2 \text{e}^{2 ( \beta_0 - \varphi )}\left(
2f \partial_\phi \beta_0- \partial_\phi f  \right)
\nonumber
\\
&&
-\text{e}^{-\varphi  }\left[ Y  \left( \partial_\phi U_0 +U_0 \partial_\phi  \varphi + \partial_u\varphi \right)+U_0  \partial_\phi Y + \partial_uY\right] ,
\\
\delta_\xi\beta_0&=& -\frac{\omega}{2}+\frac{1}{2}\text{e}^{-2  \beta_0  } \left[U_0 \left(f \partial_\phi \varphi -\partial_\phi f\right)-\partial_u f +f  \left( \partial_\phi U_0 +\partial_u\varphi \right)\right]
\nonumber
\\
&&
+\frac{1}{2}\text{e}^{-\varphi  } \left(\partial_\phi Y-2  \partial_\phi\beta_0  Y \right) ,
\end{eqnarray}
as well as
\begin{eqnarray}
\delta_\xi N&=&\omega N +\text{e}^{-\varphi  } \left(Y  N \partial_\phi  \varphi  -Y\partial_\phi N -2 N  \partial_\phi Y \right)+ \text{e}^{-2 \beta_0 } \left\{\partial_u N
\right.
\nonumber
\\
&&\left.-f  \left[U_0  \left(\partial_\phi N +N\partial_\phi   \varphi  \right)+N  \left(2 \partial_\phi U_0 +\partial_u \varphi \right)\right]
-M  \left(\partial_\phi f -2 f  \partial_\phi \beta_0 \right) \right\} 
\nonumber
\\
&&
+\text{e}^{-2 \varphi } \left\{\partial_\phi f  \left[
 2 \left(\partial_\phi \varphi\right)^2
 -\partial_\phi^2 \varphi 
 +4 \partial_\phi \beta_0 \partial_\phi  \varphi  
 -8 \left(\partial_\phi \beta_0\right)^2
 -4 \partial_\phi^2 \beta_0 \right]
\right.
\nonumber
\\
&&
+2 f  \left[2 \partial_\phi \varphi \left(\partial_\phi \beta_0\right)^2 +\partial_\phi \beta_0 \left(\partial_\phi^2 \varphi  -2 \partial_\phi^2 \beta_0 -2\left(\partial_\phi \varphi \right)^2\right)
\right.
\nonumber
\\
&&
\left.\left.
+3\partial_\phi \varphi \partial_\phi^2 \beta_0  +4 \left(\partial_\phi \beta_0\right)^3-\partial_\phi^3\beta_0 \right]-3 \partial_\phi \varphi   \partial_\phi^2 f+\partial_\phi^3 f \right\},
\end{eqnarray}
and
\begin{eqnarray}
\delta_\xi M&=& \omega M   -\text{e}^{-\varphi   }\left(  Y\partial_\phi M  +M   \partial_\phi Y  \right)  -\text{e}^{-2\beta_0 }\left\{f   \left( U_0 \partial_\phi M   +\partial_u M \right) 
\right.
\nonumber
\\
&&
\left.+M\left[U_0   \left(\partial_\phi f  +f   \left(\partial_\phi \varphi  -2   \partial_\phi \beta_0  \right)\right)
+\partial_u f  +f   \left(\partial_\phi U_0  -2  \partial_u  \beta_0  +\partial_u \varphi  \right)
\right] \right\}
\nonumber
\\
&&+2 \text{e}^{-2 \varphi   } \left\{-2 k^2  \text{e}^{2   \beta_0  } N  \left(\partial_\phi f  -2 f     \partial_\phi \beta_0  \right)-2 \partial_\phi f   \partial_\phi U_0  \partial_\phi \varphi   + \partial_\phi U_0 \partial_\phi^2 f-\partial_\phi f  \partial_\phi^2U_0 
\right.
\nonumber
\\
&&
+U_0   \left[\partial_\phi f   \left(2   \partial_\phi \beta_0  \partial_\phi \varphi   -2  \partial_\phi^2 \beta_0 -2\partial_\phi^2\varphi  -4 \left( \partial_\phi  \beta_0\right)^2+\left(\partial_\phi \varphi  \right)^2\right)-2   \partial_\phi \varphi \partial_\phi^2f   
\right.
\nonumber
\\
&&
\left.
 + \partial_\phi^3 f +2 f   \left(\left(2  \partial_\phi  \beta_0 -\partial_\phi \varphi  \right) \left(  \partial_\phi \beta_0   \partial_\phi \varphi  -2  \partial_\phi^2 \beta_0  \right)
+ \partial_\phi  \beta_0\partial_\phi^2  \varphi   -  \partial_\phi^3 \beta_0  \right)\right]
\nonumber
\\
&&
+2 \partial_u f    \partial_\phi  \beta_0   \partial_\phi \varphi  -4 \partial_u f   \left(  \partial_\phi \beta_0\right)^2-2 \partial_u f    \partial_\phi^2 \beta_0+\partial_\phi f  \partial_\phi  \varphi \partial_u  \varphi   - \partial_u \varphi \partial_\phi^2f  
\nonumber
\\
&&
- \partial_\phi \varphi \partial_u \partial_\phi f   -2 \partial_\phi f  \partial_u \partial_\phi  \varphi +\partial_u \partial_\phi^2 f +2 f    \partial_\phi U_0     \partial_\phi \beta_0  \partial_\phi \varphi  -4 f  \partial_\phi  U_0 \left( \partial_\phi   \beta_0\right) ^2
\nonumber
\\
&&
-2 f   \partial_\phi U_0 \partial_\phi^2   \beta_0+4 f   \left( \partial_\phi  \beta_0\right)^2 \partial_u \varphi   -2 f     \partial_\phi \beta_0   \partial_\phi \varphi  \partial_u  \varphi   +2 f   \partial_u  \varphi \partial_\phi^2  \beta_0 \nonumber
\\
&&
 +2 f    \partial_\phi \varphi   \partial_u \partial_\phi  \beta_0  +2 f    \partial_\phi  \beta_0  \partial_u \partial_\phi \varphi  
-8 f    \partial_\phi  \beta_0    \partial_u \partial_\phi  \beta_0 -2 f    \partial_u \partial_\phi^2 \beta_0
\nonumber
\\
&&
\left.
 +2 \text{e}^{2   \beta_0  }   \partial_\phi \beta_0\partial_\phi  \omega  +2 \text{e}^{2   \beta_0 -\varphi }  \partial_\phi  \beta_0  \left(\partial_\phi \varphi    \partial_\phi Y -\partial_\phi^2Y  \right) \right\}.
\end{eqnarray}
Again these expressions are valid for non-vanishing or vanishing $k$, i.e. for locally AdS or locally Minkowski. The corresponding algebras, however are distinct. 

\section{On the algebra of residual diffeomorphisms}
\label{app2}

The most general three-dimensional Einstein or Ricci-flat solution in Eddington--Finkelstein gauge of fluid/gravity correspondence has been worked out in Sec. \ref{genbulk}. It depends on six arbitrary functions and possesses residual diffeomorphisms, generated by the vector $\xi$ given in \eqref{kiladsgen} for anti de Sitter, and parameterized by four functions of the boundary coordinates: $f$, $Y$, $S$ and $Z$. Four auxiliary functions have also been introduced:
\begin{eqnarray}
\psi&=&S +\frac{\Theta^\ast}{k}Y+\text{u}( f),
\label{Psist}
\\
\label{Psi}
\psi^\ast&=&k
Z -\Theta^\ast f  +\ast\text{u}( f ),
\\
\omega&=&S +\Theta f  + \frac{\ast\text{u}(Y)}{k},
\label{W}
\\
k \omega^\ast&=&k^2 Z -\Theta Y +\text{u}(Y),
\label{Wst}
\end{eqnarray}
as coefficients in the variations of the fundamental fields $u_\mu$ and $\ast u_\mu$  (Eqs. \eqref{deltuastud}).

With the modified Lie bracket introduced in \eqref{modlie}, the generating vector fields form an algebra, which obeys Eqs. \eqref{AdS-algebra-f}, \eqref{AdS-algebra-Y}, \eqref{AdS-algebra-S}, \eqref{AdS-algebra-Z}. We have already pointed out that if $f_a$, $Y_a$, $S_a$, $Z_a$ are field-dependent , the last terms, like $\delta_{\xi_1}f_2-\delta_{\xi_2}f_1$ in \eqref{AdS-algebra-f}, do not vanish. This happens if we chose as fundamental parameters some of the auxiliary functions introduced in \eqref{Psist}, \eqref{Psi}, \eqref{W}, \eqref{Wst}, or any combination thereof, in which case the $f_a$, $Y_a$, $S_a$, $Z_a$ will depend on the boundary velocity fields and be sensitive to another diffeomorphism. A standard paradigm stems out of the requirement that the boundary vector $\xi_{(0)}$ \eqref{Kilbdy},  which packages conveniently two of the four functions ($f$ and $Y$) and is duplicated here for clarity
\begin{equation}
\label{Kilbdy-bis}
\xi _{(0)}=  f \text{u} +\frac{Y}{k}\ast \text{u},
\end{equation}
be a fundamental parameter i.e. field-independent. Demanding $\delta_{\xi_1}\xi_{(0)2}=0$ and using the transformation rules \eqref{deltuastuU}, we find 
\begin{equation}
\label{Kilbdy-inv}
\delta_{\xi_1}f_2=-\psi_1f_2-\psi^\ast_1\frac{Y_2}{k}, \quad 
\delta_{\xi_1}Y_2=-k\omega^\ast_1f_2-\omega_1Y_2.
\end{equation}
With those variations, Eqs. \eqref{AdS-algebra-f}, \eqref{AdS-algebra-Y} are recast as 
\begin{equation}
\begin{split}
f_3&=f_1 \text{u}\left(f_2\right)-f_2 \text{u}\left(f_1\right)+\frac{\Theta^\ast}{k}\left(f_1 Y_2 - f_2 Y_1\right)+\frac{1}{k}\left(Y_1 \ast\text{u}\left(f_2\right)-Y_2 \ast\text{u}\left(f_1\right)\right),
\\
Y_3&=
\frac{1}{k}\left(Y_1 \ast\text{u}\left(Y_2\right)-Y_2 \ast\text{u}\left(Y_1\right)\right)+\Theta\left(Y_1 f_2   -Y_2f_1\right)+f_1 \text{u}\left(Y_2\right)-f_2 \text{u}\left(Y_1\right)
,
\end{split}
\label{AdS-algebra-fy}
\end{equation}
which are equivalent to  
\begin{equation}
\label{AdS-algebra-xi0}
\xi_{(0)3}
=\left[\xi_{(0)1},\xi_{(0)2}\right].
\end{equation}
Equations  \eqref{AdS-algebra-S}, \eqref{AdS-algebra-Z} can also be expressed using the boundary vector fields $\xi_{(0)a}$:
\begin{eqnarray}
S_3&=&\xi_{(0)1} \left(S_2\right)-\xi_{(0)2} \left(S_1\right)+\delta_{\xi_1}S_2-\delta_{\xi_2}S_1,
\label{algebra-S}
\\
Z_3&=&\xi_{(0)1} \left(Z_2\right)-\xi_{(0)2} \left(Z_1\right)+\delta_{\xi_1}Z_2-\delta_{\xi_2}Z_1.
\label{algebra-Z}
\end{eqnarray}

The general structure of the algebra is a semi-direct product of the $\{f,Y\}$ component with two extra factors, $S$ and $Z$. The final form of the latter depends on the variations of $S$ and $Z$. 
\begin{itemize}
\item If $S$ and/or $Z$ are chosen as fundamental parameters the last two terms in Eqs. 
\eqref{algebra-S} and/or \eqref{algebra-Z} drop: $\delta_{\xi_1}S_2=\delta_{\xi_2}S_1=0$ and/or 
$\delta_{\xi_1}Z_2=\delta_{\xi_2}Z_1=0$.

\item It may be convenient to select alternatively $\psi$ and/or $\psi^\ast$ as fundamental parameters, and thus demand 
$\delta_{\xi_1}\psi_2=\delta_{\xi_2}\psi_1=0$ and/or $\delta_{\xi_1}\psi^\ast_2=\delta_{\xi_2}\psi^\ast_1=0$, which, using \eqref{Psist}, \eqref{Psi}, \eqref{W}, \eqref{Wst}, make it possible to determine  $\delta_{\xi_1}S_2$, $\delta_{\xi_2}S_1$, $\delta_{\xi_1}Z_2$ and $\delta_{\xi_2}Z_1$. Putting everything together, we find 
\begin{eqnarray}
\psi_3&=&\psi^\ast_1 \omega^\ast_2-\psi^\ast_2 \omega^\ast_1
\nonumber
\\
&=&\frac{\psi^\ast_1}{k}
\left(\text{u}\left(Y_2\right)-\Theta Y_2 -k\ast\text{u}\left( f_2 \right)+k\Theta^\ast f_2  
\right)
\nonumber
\\
&&-\frac{\psi^\ast_2}{k}
\left(\text{u}\left(Y_1\right)-\Theta Y_1  -k\ast\text{u}\left( f_1 \right)+k\Theta^\ast f_1 
\right),
\label{algebra-psist}
\end{eqnarray}
and/or
\begin{eqnarray}
\psi^\ast_3&=&\psi^\ast_1 \left(\omega_2-\psi_2\right)-\psi^\ast_2  \left(\omega_1-\psi_1\right)
\nonumber
\\
&=&\frac{\psi^\ast_1}{k}
\left(\ast\text{u}\left(Y_2\right)-\Theta^\ast Y_2  -k\text{u}\left( f_2 \right)+k\Theta f_2 
\right)
\nonumber
\\
&&-\frac{\psi^\ast_2}{k}
\left(\ast\text{u}\left(Y_1\right)-\Theta^\ast Y_1-k\text{u}\left( f_1 \right) +k\Theta f_1  
\right),
\label{algebra-psi}
\end{eqnarray}
showing in passing that the directions $\psi$ and $\psi^\ast$ are inequivalent inside the algebra.
\end{itemize}
Possible combinations defining the full algebra, besides $\xi_{(0)}$, are $\{S,Z\}$,  $\{\psi,Z\}$,  $\{S,\psi^\ast\}$ or  $\{\psi,\psi^\ast\}$.

The same pattern can be pursued for locally Minkowski three-dimensional spacetimes. Again four functions $f$, $Y$, $S$ and $Z$ capture all information about the asymptotic Killing vectors \eqref{kiladsgencar}, which now obey \eqref{flat-algebra-f}, \eqref{flat-algebra-Y}, \eqref{flat-algebra-S} and \eqref{flat-algebra-Z}. The boundary vector \eqref{Kilbdy-bis} subsists as a tangent field over the Carrollian boundary spacetime
\begin{equation}
\label{Kilcarbdy-bis}
\xi _{(0)}=  f \upsilon +Y\ast \upsilon,
\end{equation}
and requiring its invariance as $\delta_{\xi_1}\xi_{(0)2}=0$, is equivalent to 
\begin{equation}
\label{Kilcarbdy-inv}
\delta_{\xi_1}f_2=-\psi_1f_2-\alpha_1Y_2, \quad 
\delta_{\xi_1}Y_2=- \left(\upsilon(Y)-\theta Y
\right)f_2-\omega_1Y_2.
\end{equation}
Observe that the auxiliary functions $\psi$ and $\omega$, defined in \eqref{Psist} and \eqref{W}, are well-defined in the Carrollian limit, as is $\alpha=\nicefrac{\psi^\ast}{k}$ in  \eqref{Psi}:
\begin{eqnarray}
\psi&=&S +\theta^\ast Y+\upsilon( f),
\label{PsistC}
\\
\label{PsiC}
\alpha&=&
Z -\theta^\ast f  +\ast\upsilon( f ),
\\
\omega&=&S +\theta f  + \ast\upsilon(Y).
\label{WC}
\end{eqnarray}
However, $k\omega^\ast$ in \eqref{Wst} 
ceases depending on $Z$: its vanishing-$k$ limit is $\upsilon(Y)-\theta Y$, and this is at the heart of the breakup between flat and anti-de Sitter asymptotic symmetry algebras.  Equations \eqref{flat-algebra-f}, \eqref{flat-algebra-Y} now read:
\begin{equation}
\begin{split}
f_3&=f_1  \upsilon\left(f_2\right)-f_2 \upsilon\left(f_1\right)+\theta^\ast\left(f_1 Y_2 - f_2 Y_1\right)+Y_1 \ast \upsilon\left(f_2\right)-Y_2 \ast \upsilon\left(f_1\right),
\\
Y_3&=
Y_1 \ast \upsilon\left(Y_2\right)-Y_2 \ast \upsilon\left(Y_1\right)+f_1\left( \upsilon\left(Y_2\right)-\theta Y_2\right)-f_2\left(\upsilon\left(Y_1\right)-\theta Y_1\right),
\end{split}
\label{flat-algebra-fy}
\end{equation}
so that \eqref{AdS-algebra-xi0} still holds for the vectors \eqref{Kilcarbdy-bis}, and the full algebra includes \eqref{algebra-S} and  \eqref{algebra-Z}. Again, we can chose 
$\delta_{\xi_1}\alpha_2=\delta_{\xi_2}\alpha_1=0$ and/or $\delta_{\xi_1}\psi_2=\delta_{\xi_2}\psi_1=0$,
and express \eqref{algebra-S} and  \eqref{algebra-Z} for $\psi$ and/or $\alpha$ using \eqref{PsiC}:
\begin{eqnarray}
\psi_3&=&\alpha_1
\left(\upsilon\left(Y_2\right)-\theta Y_2 
\right)-\alpha_2
\left(\upsilon\left(Y_1\right)-\theta Y_1
\right),
\label{algebra-psistC}
\\
\alpha_3&=&\alpha_1\left(\omega_2-\psi_2\right)
-\alpha_2\left(\omega_1-\psi_1\right)
\nonumber
\\
\nonumber
&=&\alpha_1
\left(\ast\upsilon\left(Y_2\right)-\theta^\ast Y_2 -\upsilon\left( f_2 \right) +\theta f_2 
\right)
\nonumber
\\
&&-\alpha_2
\left(\ast\upsilon\left(Y_1\right)-\theta^\ast Y_1  -\upsilon\left( f_1 \right)+\theta f_1 
\right).
\label{algebra-alpha}
\end{eqnarray}
The option still exists to adopt $S$ and/or $Z$ as fundamental parameters, in which case Eqs. \eqref{algebra-S} and/or \eqref{algebra-Z} are in use, after discarding the last two terms.

At first glance, the asymptotic symmetries of three-dimensional locally flat or anti-de Sitter spacetimes seem very similar (e.g. Eqs. \eqref{AdS-algebra-fy} vs. \eqref{flat-algebra-fy}). However, significant differences turn up when investigating specific corners of the above algebras, as we perceive for instance in comparing \eqref{algebra-psist} and \eqref{algebra-psistC}. This deserves a separate comprehensive study, out of place here. We can nonetheless rapidly browse through a few illustrations.
\begin{enumerate}

\item \textbf{Boundary Weyl transfromations.} The variation of the boundary metric under general residual diffeomorphisms of anti-de Sitter bulk solutions is captured in Eq. \eqref{delatxids2}:
\begin{equation}
\delta_\xi \text{d}s^2=\frac{2}{k^2}\left(\psi\mathbf{u}^2
-\omega\ast\mathbf{u}^2
+\left(\omega^\ast -\psi^\ast
\right)\mathbf{u} \ast\mathbf{u}\right).
\label{delatxids2-bis}
\end{equation}
We infer from the latter that a diffeomorphism generates a boundary Weyl transformation under the conditions  $\psi=\omega$ and $\psi^\ast=\omega^\ast$, which explicitly read (using \eqref{Psist}, \eqref{Psi}, \eqref{W} and \eqref{Wst}):
\begin{equation}
\label{fYWeyl}
\text{u}( f)+\frac{\Theta^\ast}{k}Y= \frac{\ast\text{u}(Y)}{k}+\Theta f ,
\quad
\text{u}(Y)-\Theta Y = k\ast\text{u}( f )-k\Theta^\ast f .
\end{equation}
These equations are equivalent to demanding that $\xi_{(0)}$ be a boundary conformal Killing field.\footnote{Setting  $\gamma=\Gamma=1$ and $\Delta=v^\phi=0$ in the parameterization \eqref{genf}, conditions \eqref{fYWeyl} translate into the usual chirality requirements: 
$\partial_u f=\partial_\phi Y$ and $\partial_u Y=k^2\partial_\phi f$.} 
Indeed, the Lie derivative of $\text{d}s^2$ along $\xi_{(0)}$ evaluates to\footnote{We use  result
$\left[\xi_{(0)},\text{u}\right]=\left(S-\psi\right)\text{u}+\left(kZ-\omega^\ast\right)\ast\text{u}$
and
$\left[\xi_{(0)},\ast\text{u}\right]=\left(kZ-\psi^\ast\right)\text{u}+\left(S-\omega\right)\ast\text{u}$, which has a Carrollian counterpart: $\left[\xi_{(0)},\upsilon\right]=\left(S-\psi\right)\upsilon+\left(\theta Y-\upsilon(Y)\right)\ast\upsilon$
and
$\left[\xi_{(0)},\ast\upsilon\right]=\left(Z-\alpha\right)\upsilon+\left(S-\omega\right)\ast\upsilon$.} 
\begin{equation}
\label{Liegbry}
\mathscr{L}_{\xi_{(0)}}\text{d}s^2
=\frac{2}{k^2}
\left(-\left(\psi-S\right)\mathbf{u}^2+ \left(\omega-S\right)\ast\mathbf{u}^2+ \left(\psi^\ast-\omega^\ast\right)
\mathbf{u}\ast\mathbf{u}\right),
\end{equation}
which under the above Weyl conditions, simplifies to 
$\mathscr{L}_{\xi_{(0)}}\text{d}s^2=2\left(\omega-S\right) \text{d}s^2$.
Hence, the diffeomorphisms at hand are still characterized by four functions $f$, $Y$, $\psi$ and $\psi^\ast$ (or $S$ and $Z$), although the former two are not arbitrary, but must obey \eqref{fYWeyl}. Their algebra is the product of the boundary conformal algebra (satisfying \eqref{AdS-algebra-xi0}) with two extra factors (obeying \eqref{algebra-psist} and \eqref{algebra-psi}): the boundary local Lorentz boosts $\psi^\ast$ -- hydrodynamic-frame transformations -- and the boundary rescalings $\psi$. 
Both are abelian ideals due to the Weyl conditions: $
\psi_3=0$ and $\psi^\ast_3=0$.
The result for $\psi$ is in line with the analysis of  \cite{Barnich:2010eb}. We have more here with 
$\psi^\ast$ -- vanishing in Bondi gauge. 

In locally flat bulk spacetimes, residual diffeomorphisms produce variations of the form \eqref{dl2lim} on the Carrollian degenerate boundary metric, 
\begin{equation}
\label{dl2lim-bis} 
\delta_\xi \text{d}\ell^2 = 2\ast \pmb{\mu}\delta_\xi \ast\pmb{\mu}=-2 \omega\text{d}\ell^2 
+2\left(\upsilon(Y)-\theta Y
\right)\pmb{\mu}\ast \pmb{\mu}
,
\end{equation}
which are Weyl transformations if 
\begin{equation}
\label{fYWeylC}
\upsilon(Y)=
\theta Y
\end{equation}
with no further restriction on $\omega$ or $\psi$. This is a generalization of the more familiar requirement $\partial_u Y=0$, reached for $\upsilon=\partial_u $ (i.e. $\gamma=\Gamma=1$, $\Delta=v^\phi=0$ in the parameterization \eqref{genfvc}), which confers $Y$ the status of circle-diffeomorphism (or line, more generally) generator. Condition \eqref{fYWeylC} simplifies the $Y$ component in \eqref{flat-algebra-fy}:  
\begin{equation}
Y_3=Y_1 \ast \upsilon\left(Y_2\right)-Y_2 \ast \upsilon\left(Y_1\right),
\label{flat-algebra-fy-con}
\end{equation}
giving the $\{f,Y\}$ algebra the standard semi-direct-product structure of conformal Carroll groups (see e.g. \cite{Duval:2014uva, Duval:2014lpa}). The imprint of the conformal nature of the symmetries at hand is also visible in the Lie derivative of the Carrollian metric along $\xi_{(0)}$:
\begin{equation}
\label{LiegbryC}
\mathscr{L}_{\xi_{(0)}}\text{d}\ell^2
=2
 \left(\omega-S\right)
 \text{d}\ell^2+ 2\left(
 \upsilon(Y)-
\theta Y
 \right)
\pmb{\mu}\ast\pmb{\mu}.
\end{equation}
Summarizing, the residual diffeomorphisms of flat spacetimes that induce a Weyl transformation on the Carrollian boundary are described in terms of  four functions $f$, $Y$, $\psi$ and $\alpha$ (or $S$ and $Z$) subject to the constraint \eqref{fYWeylC}. The algebra has three (semi-)direct factors, associated with  $Y$, $\psi$ and $\alpha$ (or $S$ and $Z$). Amongst the last two, the former, corresponding to boundary rescalings, is abelian under the requirement of induced boundary Weyl transformations because $\psi_3=0$ (see \eqref{algebra-psistC}), in agreement with \cite{Barnich:2010eb, Barnich:2009se}. However, according to \eqref{algebra-alpha}, $\alpha_3\neq 0$ so that local Carroll boosts do not define an abelian ideal; this should be contrasted with the analogue anti-de Sitter situation. 

Bulk diffeomorphisms leaving the boundary metric invariant are the subset $\omega=0$ of those under consideration here. These include the Lorentz or Carroll boosts ($\psi^\ast$ or $\alpha$) studied in more detail in Sec. \ref{genbulk}, together with the boundary diffeomorphisms generated by conformal Killing vector fields ($f$ and $Y$). For locally Minkowski solutions, the fourth direction $\psi$ remains unaffected, while it drops for anti de Sitter.

\item \textbf{Locking the hydrodynamic frame.} The function  $\psi^\ast$ (or equivalently $Z$) controls the diffeomorphisms that produce a change in the boundary hydrodynamic frame -- see Sec. \ref{genbulk}. Firming-up  $\psi^\ast$ disables those transformations. This occurs in the Bondi gauge, and can be generalized by keeping $u_\phi$ or $\mu_\phi$ non-zero but fixed, as opposed to Bondi, where $u_\phi=0$ or $\mu_\phi=0$, realized with $\Delta=0$. Hence,  imposing $\delta_\xi u_\nu=0$ or $\delta_\xi \mu_\nu=0$ in \eqref{deltuastud} or  \eqref{deltaudexpllim} with $\Delta\neq0$, we obtain a generalization of  \eqref{ZfY}:
\begin{equation}
\label{psipsistalphfYDelta}
\psi^\ast=k\frac{\Delta}{\Gamma}\psi
\quad\text{or}\quad
\alpha=\frac{\Delta}{\Gamma}\psi.
\end{equation}

The residual diffeomorphisms are parameterized by $f$, $Y$ and $\psi$. For anti de Sitter, these diffeomorphisms obey the algebra \eqref{AdS-algebra-xi0} and \eqref{algebra-psist}, while \eqref{algebra-psi} drops off:
\begin{eqnarray}
\psi_3&=&\frac{\Delta}{\Gamma}\left[\frac{\psi_1}{k}
\left(\text{u}\left(Y_2\right)-\Theta Y_2 -k\ast\text{u}\left( f_2 \right)+k\Theta^\ast f_2  
\right)\right.
\nonumber
\\
&&-\left.\frac{\psi_2}{k}
\left(\text{u}\left(Y_1\right)-\Theta Y_1  -k\ast\text{u}\left( f_1 \right) +k\Theta^\ast f_1
\right)\right].
\label{algebra-psist-genbon}
\end{eqnarray}
For locally Minkowski spacetimes the algebra is   \eqref{flat-algebra-fy} and \eqref{algebra-psistC}: 
 \begin{equation}
\psi_3=\frac{\Delta}{\Gamma}\left[\psi_1
\left(\upsilon\left(Y_2\right)-\theta Y_2 
\right)-\psi_2
\left(\upsilon\left(Y_1\right)-\theta Y_1
\right)\right].
\label{algebra-psistC-genbon}
\end{equation}

In Bondi gauge, where $\Delta=0$, $\psi$ supports an abelian ideal irrespective of $k$, which completes the $\{f,Y\}$ component of the residual-diffeomorphism algebra. This enhancement of the symmetry algebra was obtained in Ref.  \cite{Troessaert:2013fma}. The abelian ideal survives for non-vanishing $\Delta=0$ if we further restrict the bulk diffeomorphisms to those producing boundary Weyl transformations (we combine then \eqref{algebra-psist-genbon} or 
\eqref{algebra-psistC-genbon} with \eqref{fYWeyl} or \eqref{fYWeylC}). However, for generic locked-frame (i.e. $\psi^\ast=0$) residual diffeomorphisms, this ideal is absent. 

\end{enumerate}
The next two situations are borrowed from Ref. \cite{CCMPS}, where they were used to uncover that different hydrodynamic frames describe fluids with distinct global properties, encoded in their gravity duals.
\begin{enumerate}
\setcounter{enumi}{2}

\item \textbf{Fluids without heat current.} These are relativistic fluids with $\chi=0$, or Carrollian fluids with $\zeta=0$. They are dual to a narrowed space of anti-de Sitter or Ricci-flat solutions with residual diffeomorphisms generically restricted to $Z=0$, in order to ensure $\delta_\xi\chi=0$ or  $\delta_\xi\zeta=0$ (see Eqs. \eqref{deltach} and
\eqref{deltachcar}, respectively). The algebra of these diffeomorphisms is spanned by $f$ and $Y$, plus $\psi$ (or $S$, as a matter of convenience).  
 
\label{fwhc}

\item \textbf{Fluids at rest.} This is an antipodal situation with respect to \ref{fwhc}, moving the degrees of freedom of the fluid from the velocity to the heat current. According to a certain interpretation of the hydrodynamic-frame invariance, this confers an alternative perspective on the same physical system. From our analysis the systems are distinguishable by the algebra of bulk conserved charges, foreseen in the set of residual diffeomorphisms.

A fluid at rest is a sort of dual to a Bondi fluid. It has $u^\phi=0$ or $\mu^\phi=0$, which in the parameterization \eqref{genv} or  \eqref{genfvc} amounts to setting $v^\phi=0$. Again, the solution space has five functions subject to fluid equations, and the residual diffeomorphisms must respect the defining condition. Using \eqref{deltuastuU} or \eqref{deltauUexpllim} we find $\omega^\ast=0$ i.e.
 \begin{equation}
k^2 Z =\Theta Y-\text{u}(Y) 
\quad
\text{or}
\quad
\upsilon(Y) =\theta Y.
\end{equation}
These conditions affect more the anti-de Sitter solutions, as the diffeomorphism algebra contains now $f$, $Y$ and $\psi^\ast$, than the Minkowski spacetimes, where all $f$, $Y$, $\alpha$ and $\psi$ (or S) remain with a restriction on $Y$ only. 

\label{pf}

\end{enumerate}
Further limitations to the solution space (and consequently on the residual diffeomorphisms) can be imposed on the situations \ref{fwhc} or  \ref{pf}, such as boundary Weyl-flatness or flatness -- making the holographic fluids in \ref{fwhc} genuinely perfect (according to \eqref{anomaly} or its Carrollian relative).

\end{document}